\documentclass[sigconf,screen]{acmart}

\acmBadgeR[https://www.acm.org/publications/policies/artifact-review-and-badging-current]{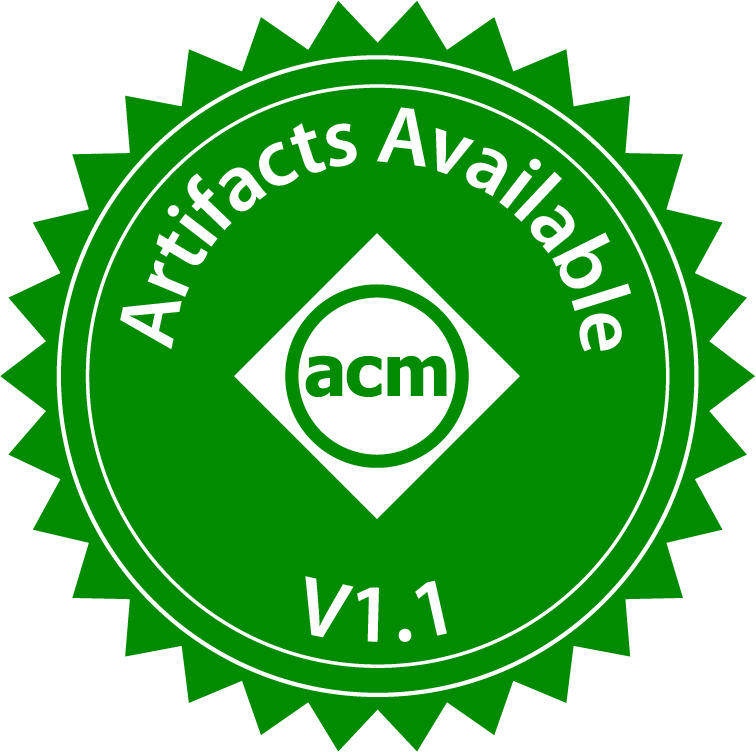}
\acmBadgeR[https://www.acm.org/publications/policies/artifact-review-and-badging-current]{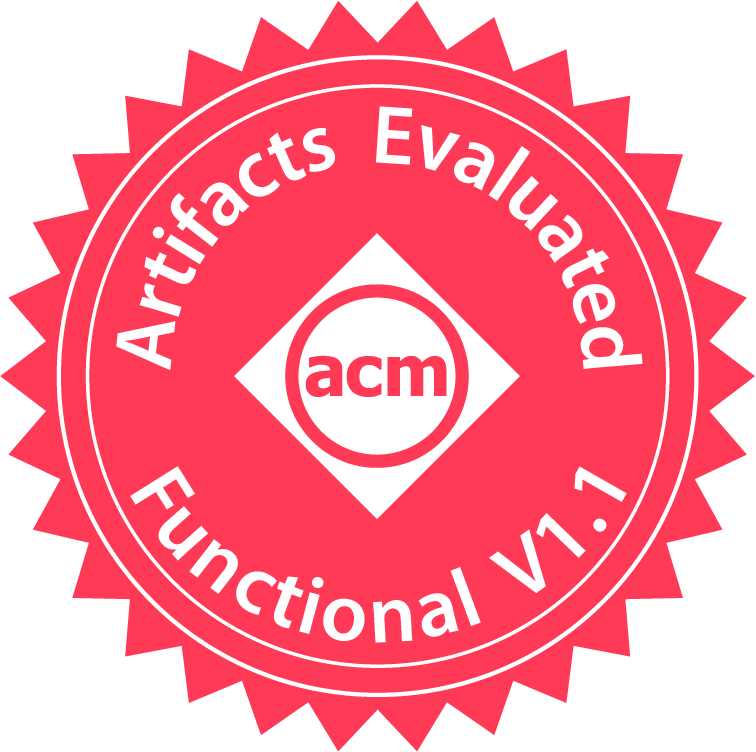}
\acmBadgeR[https://www.acm.org/publications/policies/artifact-review-and-badging-current]{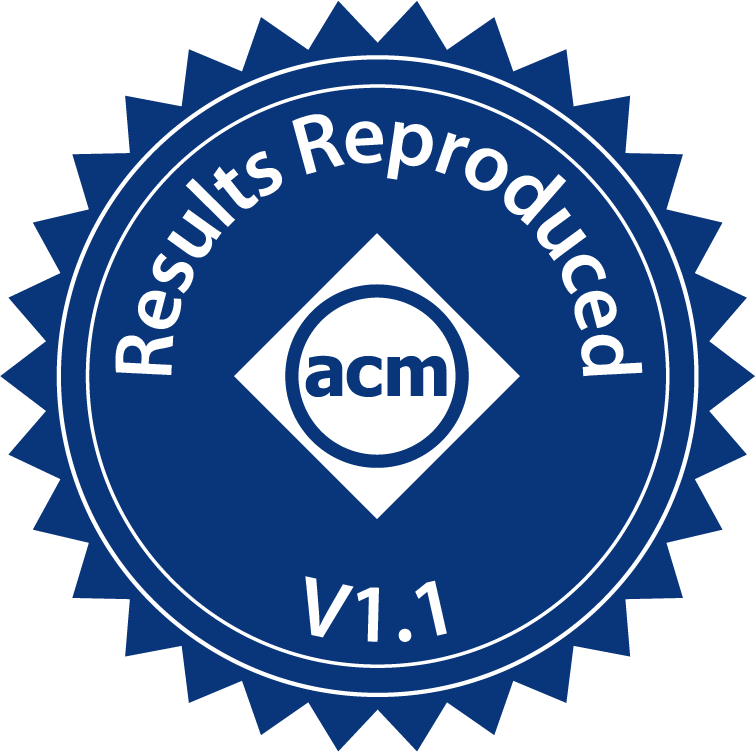}

\AtBeginDocument{%
  }

\copyrightyear{2025}
\acmYear{2025}
\setcopyright{cc}
\setcctype{by-nc-sa}
\acmConference[ISCA '25]{Proceedings of the 52nd Annual International Symposium on Computer Architecture}{June 21--25, 2025}{Tokyo, Japan}
\acmBooktitle{Proceedings of the 52nd Annual International Symposium on Computer Architecture (ISCA '25), June 21--25, 2025, Tokyo, Japan}\acmDOI{10.1145/3695053.3731015}
\acmISBN{979-8-4007-1261-6/2025/06}

\newcommand{\maa}[1]{DX100}

\newcommand{\blue}[1]{\textcolor{blue}{#1}}
\newcommand{\ignore}[1]{}

\usepackage{acmart-taps}
\usepackage{booktabs, multirow} 
\usepackage{wrapfig}
\usepackage[]{rotating}

\usepackage{listings}
\begin{document}

\title{DX100: A Programmable Data Access Accelerator for Indirection}

\author{Alireza Khadem$^{\dagger}$\textsuperscript{*}, Kamalavasan Kamalakkannan$^{\ddagger}$\textsuperscript{*}, Zhenyan Zhu$^{\dagger}$, Akash Poptani$^{\dagger}$, \\
Yufeng Gu$^{\dagger}$, Jered Benjamin Dominguez-Trujillo$^{\ddagger}$, Nishil Talati$^{\dagger}$, Daichi Fujiki\textsuperscript{\textsection}, Scott Mahlke$^{\dagger}$, \\
Galen Shipman$^{\ddagger}$, Reetuparna Das$^{\dagger}$}
\thanks{*Alireza Khadem and Kamalavasan Kamalakkannan are co-first authors.}

\affiliation{%
  \institution{$^{\dagger}$University of Michigan, $^{\ddagger}$Los Alamos National Laboratory, \textsuperscript{\textsection}Institute of Science Tokyo}
}

\email{{arkhadem,reetudas}@umich.edu}
\renewcommand{\shortauthors}{Khadem et al.}

\begin{abstract}

Indirect memory accesses frequently appear in applications where memory bandwidth is a critical bottleneck.
Prior indirect memory access proposals, such as indirect prefetchers, runahead execution, fetchers, and decoupled access/execute architectures, primarily focus on improving memory access latency by loading data ahead of computation but still rely on the DRAM controllers to reorder memory requests and enhance memory bandwidth utilization.
DRAM controllers have limited visibility to future memory accesses due to the small capacity of request buffers and the restricted memory-level parallelism of conventional core and memory systems.

We introduce \maa{}, a programmable data access accelerator for indirect memory accesses.
\maa{} is shared across cores to offload bulk indirect memory accesses and associated address calculation operations.
\maa{} \textit{reorders}, \textit{interleaves}, and \textit{coalesces} memory requests to improve DRAM row-buffer hit rate and memory bandwidth utilization.
\maa{} provides a general-purpose ISA to support diverse access types, loop patterns, conditional accesses, and address calculations. 
To support this accelerator without significant programming efforts, we discuss a set of MLIR compiler passes that automatically transform legacy code to utilize \maa{}.
Experimental evaluations on 12 benchmarks spanning scientific computing, database, and graph applications show that \maa{} achieves performance improvements of $2.6\times$ over a multicore baseline and $2.0\times$ over the state-of-the-art indirect prefetcher.

\end{abstract}





\settopmatter{printfolios=true,printacmref=false}
\sloppy
\maketitle

\section{Introduction}

Scientific and engineering simulations are critical tools used in nuclear security, materials research, energy production, aerospace, and the basic sciences~\cite{shipman2022future}. Many of these simulations routinely require weeks or even months to complete on the largest-scale supercomputers in the world, primarily bottlenecked by \textit{data access} with very low arithmetic intensity.  
For instance, state-of-the-art Adaptive Mesh Refinement (AMR) simulations~\cite{AMRIntro,rage} demand $\approx$2 PB of memory and heavily rely on sparse data structures, with over 50\% of instructions involving \textit{indirect memory accesses}~\cite{shipman2022assessing}. 
In fact, the first goal of DoE’s next-generation supercomputer, ATS-5, to be commissioned in 2027, is ``Overcoming the memory wall: continued memory bandwidth performance improvements for tri-lab applications''~\cite{ats5}.


Beyond high-performance computing, indirect memory accesses, where memory address depends on another load (\textit{e.g.}, \texttt{A[B[i]]}), are common in several data-intensive workloads.
These accesses are employed in sparse linear algebra~\cite{AMG2023,hypre} and machine learning applications~\cite{han2016eie} to use sparse data structures such as CSR and CSC, graph analytics~\cite{BC1,scottBFS,sssp,prAlgo,SHILOACH198257} to traverse graph nodes and edges, and in-memory database applications~\cite{barber2014memory, hashjoin} to join tables using hash-join algorithms.

Memory bandwidth is crucial for the performance of these applications, as indirect accesses often miss caches due to low spatial and temporal locality.
Additionally, these accesses exhibit poor memory bandwidth utilization on multi-core systems for several reasons: \textbf{First,} subsequent indirect loads often \textit{access non-contiguous memory locations} in different DRAM rows. Further, concurrent memory requests from multiple cores to different rows of the same bank lead to row conflicts. Therefore, the DRAM row-buffer hit rate drops significantly.
\textbf{Second,} \textit{memory-level parallelism (MLP) is constrained} by structural limitations within the core and memory system~\cite{8675225}, such as ROB, LSQ, and cache MSHRs.
Additionally, indirect accesses depend on prior memory accesses and address calculation instructions, creating a chain of dependencies and limiting outstanding loads and stores in the core~\cite{droplet}. As a result, DRAM controllers have limited visibility to future accesses, hindering their ability to optimize the order of DRAM commands and improve bandwidth utilization~\cite{virtualchannel,frfcfs,mutlu2008parallelism,mcrl,streammc,historymc,brustmc}. Previous approaches such as runahead execution~\cite{runahead_org,runahead_ooo,runahead_efficient,emc} enhance the memory access latency by improving cache hit rate but remain bounded by core's structural limitations.

The research community has focused extensively on compute acceleration~\cite{eyeriss,TeraOPS,pedal,scnn,gendp,TPU,HAMR,sparsetpu,physicsacc,fastmotifs,cambricon,han2016eie,bitset}, but less attention has been paid to building accelerator architectures for data access~\cite{spzip, lee2024terminus, wang2024data}. Data-centric accelerator solutions can enrich future computing systems especially when memory access and data movement costs dominate~\cite{6757323}. \textbf{This work proposes \maa{}, a programmable data access accelerator that enables the offloading of \textit{bulk} indirect loads, stores, and read-modify-write (RMW) operations.} Accelerating bulk memory accesses using \maa{} significantly reduces the core’s instruction footprint. More importantly, to enhance the memory bandwidth utilization, \maa{} is inspired by the following key ideas: (a) Offloading indirect memory accesses and address calculation instructions to an accelerator near the memory controllers improves the effective memory-level parallelism. (b) Indirect address issue to DRAM can be optimized for \textit{bulk} memory operations by leveraging the visibility of a large window of indices. Consider a bulk access \texttt{A[B[i]]} for \texttt{i} ranging from 0 to 16K. \maa{} has the visibility of all 16K indices after fetching them. So, \maa{} \textit{reorders} them to improve the row-buffer hit rate, \textit{coalesces} them to reduce accesses, and \textit{interleaves} them to enhance DRAM channel and bank-group interleaving.

\maa{} provides a general-purpose ISA with eight instructions to support a diverse set of loop patterns (\textit{single} and \textit{range}), loop condition patterns, multiple levels of indirection, address calculation operations (\textit{scalar} and \textit{vector}), and access types (\textit{load}, \textit{store}, and \textit{RMW}).
To enable these instructions, \maa{} integrates four functional units:
(a) \textit{Stream} unit handles streaming loads and stores,
(b) \textit{Indirect} unit manages indirect loads, stores, and RMWs while facilitating reordering, coalescing, and interleaving for improving memory bandwidth utilization,
(c) \textit{Range Fuser} unit supports merging range loops to enable bulk access,
and (d) \textit{ALU} unit performs comparison and arithmetic operations.
A scratchpad unit stores the intermediate data and facilitates the communication between \maa{}'s functional units and the CPU cores. While \maa{} provides a flexible ISA, re-developing legacy code with these APIs could be cumbersome. Therefore, we develop a compiler infrastructure using the MLIR framework~\cite{lattner2021mlir} and Polygeist~\cite{moses2021polygeist} to automatically detect arbitrary indirect memory accesses within legacy code, hoist them, and transform them into \maa{} instructions.

Similar to \maa{}, Fetcher units~\cite{spzip,lee2024terminus} and Decoupled Access Execute (DAE) architectures~\cite{dae} hoist memory access instruction streams and offload them to tightly-coupled access coprocessors, improving the core structural limitations. Distinct from these works, \maa{} is shared among cores and directly accesses the main memory, eliminating the bottleneck of core and memory system structures. More importantly, these solutions remain bounded by the random memory access patterns as they do not \textit{reorder} and \textit{interleave} memory accesses across a tile of bulk indices. Consequently, they provide modest improvements in memory bandwidth utilization primarily by increasing the memory access rate. 


\ignore{
While \maa{} provides a flexible ISA, re-developing legacy code with these APIs could be cumbersome. Therefore, we develop a compiler infrastructure using the MLIR framework~\cite{lattner2021mlir} and Polygeist~\cite{moses2021polygeist} to automatically detect arbitrary indirect memory accesses within legacy code and transform them into \maa{} instructions.
Specifically, the compiler first tiles loops based on hardware-specific tile sizes and performs a depth-first search starting from loop induction variables to identify indirect memory access patterns.
Once detected, these accesses are mapped to external \maa{} library functions, enabling seamless offloading to the accelerator without extensive code modifications.
}

In summary, this paper offers the following contributions:

\begin{itemize}

\item We propose \maa{}, a programmable data access accelerator that supports offloading indirect and streaming memory accesses. \maa{} enhances memory bandwidth utilization by optimizing \textit{bulk} accesses: it \textit{reorders} to improve DRAM row-buffer hits, \textit{coalesces} to reduce accesses, and \textit{interleaves} to enhance DRAM channel and bank-group parallelism. 

\item The proposed architecture and ISA are designed for integration with arbitrary applications that can benefit from bulk memory accesses. \maa{} is flexible to support multiple degrees of indirection, diverse sets of loop patterns, loop conditions, data access types (load, store, RMW), and various address calculation operations.

\item We accurately evaluate \maa{} as a memory-mapped accelerator using execution-based and event-driven simulators~\cite{gem5, ramulator2}. We study several benchmark suites, such as NAS~\cite{nas}, GAP~\cite{gap}, Hash-Join~\cite{hashjoin}, UME~\cite{ume}, and Spatter~\cite{spatter}. \maa{} achieves a $2.6\times$ performance improvement over the multicore baseline and outperforms the state-of-the-art hardware indirect prefetcher, DMP~\cite{dmp}, by $2.0\times$.

\end{itemize}

\ignore{
We evaluate \maa{} by integration as a memory-mapped accelerator within the execution-based and cycle-accurate GEM5 simulator~\cite{gem5}, utilizing Ramulator2 as the back-end memory simulator~\cite{ramulator2}.
We demonstrate \maa{} performance improvements across several benchmark suites, such as NAS~\cite{nas}, GAP~\cite{gap}, Hash-Join~\cite{hashjoin}, and Spatter~\cite{spatter}.
\maa{} achieves a $X\times$ performance improvement over the baseline and outperforms the state-of-the-art indirect prefetcher DMP~\cite{dmp} by $X\times$.
Additionally, we use microbenchmarks to study \maa{} benefits for different access patterns and cache/DRAM row-buffer hit rates.
}





\section{Background and Motivation}

\subsection{DRAM Organization and Access Reordering}

\begin{figure}[h]
    \centering
    \vspace{-2mm}
    \includegraphics[width=0.90\linewidth]{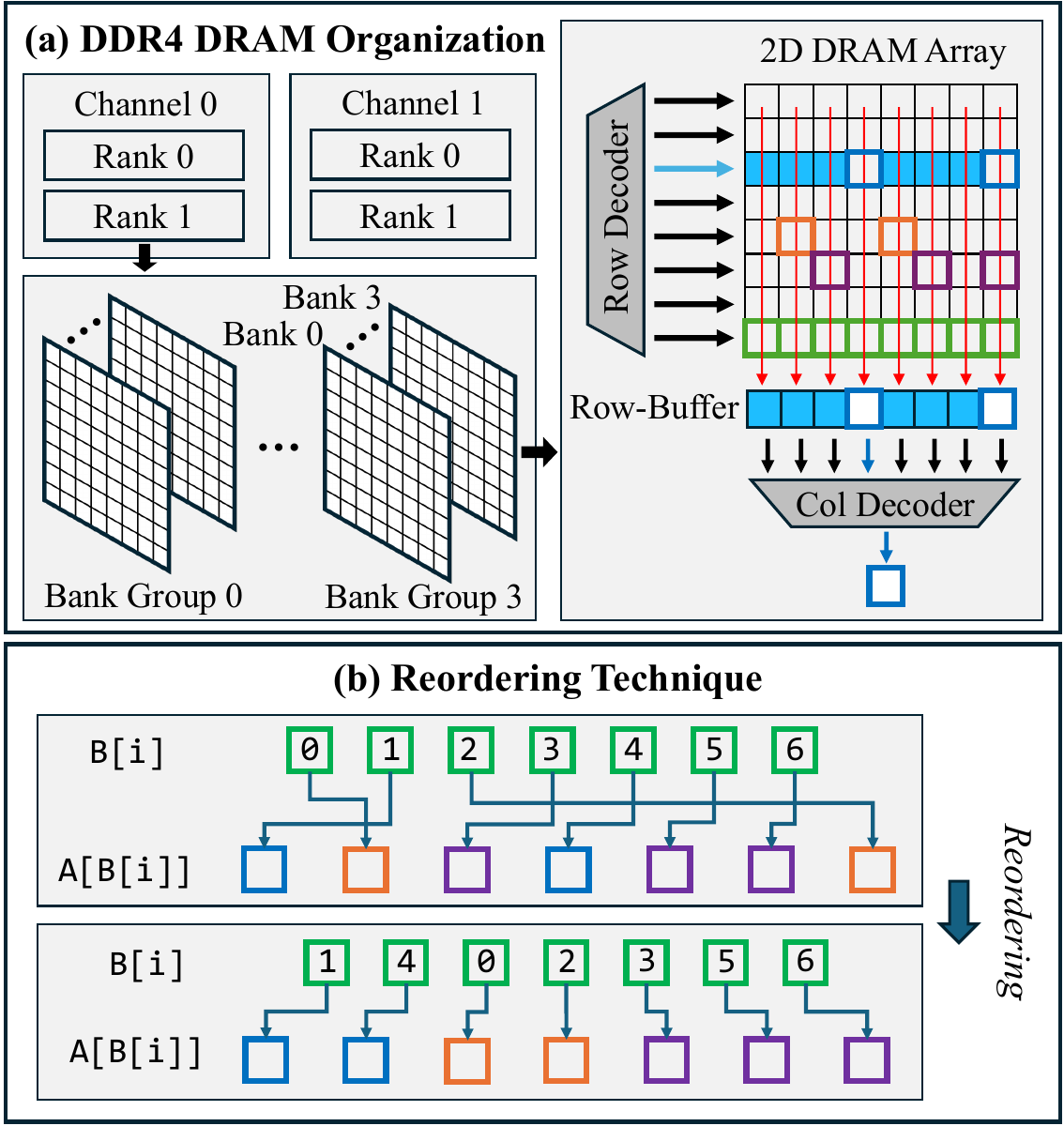}
    \vspace{-2mm}
    \caption{(a) DDR4 DRAM organization, (b) memory access reordering technique for improving row buffer hit rate.}
    \Description[DDR4 DRAM Organization and our reordering technique.]{(a) DDR4 DRAM is composed of channel, rank, bank-group, bank, rows, and columns of DRAM cells. (b) The reordering technique reorders the accesses to enhance the row-buffer hit-rate, and channel and bank-group interleaving.}
    \vspace{-2mm}
    \label{fig:access_reordering}
\end{figure}

Memory access patterns play a crucial role in bandwidth utilization due to the organization and timing constraints of DRAM.
Figure~\ref{fig:access_reordering} (a) illustrates the DRAM organization of DDR4 technology.
Memory addresses are distributed across multiple channels, ranks, four bank groups, and four banks.
Each bank consists of 2D DRAM arrays arranged in rows and columns of DRAM cells.
This hierarchical organization allows channels, ranks, bank groups, and banks to process different DRAM requests independently.
Such parallelism is essential for fully utilizing DRAM bandwidth.

Accessing data in a DRAM bank requires opening the corresponding row. This is done with a precharge (\texttt{PRE}) command to close the currently open row and then an activate (\texttt{ACT}) command to open the desired row. Once open, the data is loaded into the Row Buffer, which enables fast column accesses to the same row. Indirect memory accesses often span multiple rows, missing the row buffer and requiring frequent row switchings. Row buffer misses substantially reduce DRAM bandwidth utilization due to the long timing constraints imposed between the \texttt{PRE} and \texttt{ACT} commands.

To improve the row buffer hit rate, memory controllers often \textit{reorder} the outstanding memory accesses as shown in Figure~\ref{fig:access_reordering} (b). Both accesses targeting the blue row are issued first (A[B[1], A[B[4]). Typically, this reordering is constrained to a window of 32 to 128 accesses to balance hardware cost and complexity. This window size is not enough for applications with indirect memory access and higher sparsity, leading to lower bandwidth utilization. \maa{} architecture and programming model provide memory-bound applications the opportunity to offload bulk memory accesses to an accelerator, \textit{increasing the reordering window} to up to 16K accesses.

Even with ideal row buffer hit rates, interleaving accesses across channels and bank groups is essential for optimal DRAM bandwidth utilization. Channel interleaving is crucial because channels operate independently and can handle memory requests at a high rate.
Furthermore, consecutive accesses to the same bank group reduce bandwidth utilization due to a long column-to-column timing constraint $t_{CCDL}$.
To fully exploit the bandwidth, subsequent memory requests should be interleaved across different bank groups, reducing the effective column-to-column timing constraint to $t_{CCDS}$, which is half of $t_{CCDL}$.
\maa{} also interleaves the accesses to different channels and bank groups, effectively optimizing the utilized channel bandwidth.

\subsection{Other Challenges of Indirect Accesses}
Another key challenge for high memory bandwidth utilization is the memory request rate. For instance, under the ideal assumptions of a 100\% cache hit rate, 100\% DRAM row-buffer hit rate, and perfect interleaving, fully utilizing our evaluated system with 204.8 GB/s Last Level Cache (LLC) bandwidth and 51.2 GB/s DDR4 memory bandwidth requires one LLC request per cycle and one DRAM request every four cycles. However, conventional microarchitectures struggle to reach this rate due to several limiting factors.

First, indirect memory accesses rely on prior index loads and address calculations, forming a \textit{chain of instruction dependencies} that restricts the number of outstanding memory accesses~\cite{droplet}. \maa{} breaks this dependency chain by hoisting or sinking bulk memory accesses before or after the loop and issuing bulk accesses independently. Moreover, address calculations add substantial overhead to the \textit{dynamic instruction count}, resulting in significant execution time and energy consumption~\cite{shipman2022assessing}. \maa{} addresses this by employing custom address calculation components to accelerate bulk accesses and significantly reduce core instruction count.

Second, conventional microarchitecture contains a \textit{hierarchy of buffers} that further restricts memory-level parallelism, such as the Reorder Buffer and Load/Store Queues in the processor core, as well as the Miss Status Holding Registers (MSHRs) in the cache. \maa{} eliminates this bottleneck by bypassing intermediate buffers and directly injecting DRAM requests into the memory controller, while taking the coherency protocol into consideration.

Finally, multi-threaded execution of Read-Modify-Write (RMW) operations on conventional multi-core systems requires \textit{fine-grained atomicity}, which further constrains memory-level parallelism. Atomic RMW operations use memory fences that serialize memory requests~\cite{free_atomics} and rely on cacheline or bus locking mechanisms~\cite{intel_software_manual}. In contrast, \maa{} is shared across a group of cores and maintains exclusive write access to indirect memory regions, eliminating the need for fine-grained atomic operations.

\ignore{
Modern multi-core processors enhance memory performance with techniques like out-of-order execution, multi-level cache hierarchies, and prefetchers. While streaming memory access performs well, indirect memory access often struggles due to poor prefetcher accuracy, leading to frequent cache misses. Programmable prefetchers~\cite{bfs_prefetch,event_trigerred_prefetcher,prodigy,dmp,imp} improve accuracy but are limited by complex addressing patterns and conditional accesses. More importantly, while prefetchers reduce memory latency, they do not intentionally improve memory bandwidth.

Indirect access also reduces spatial locality, reducing cache line utilization and increasing cache misses due to capacity constraints. These capacity misses introduce higher memory latency, which processors attempt to mitigate by issuing more outstanding requests -- an approach limited by load/store queues, register files, and Miss Status Holding Registers (MSHRs). Further, the Network-on-Chip (NoC) adds latency and bandwidth constraints, emphasizing the need for accelerators closer to memory.

Indirect access also requires additional instructions for index computation~\cite{shipman2022assessing} and multiple dependent loads, with performance limited by factors like the Reorder-Buffer (ROB) size and instruction throughput. Read-Modify-Write (RMW) operations on indirect structures are particularly costly due to serialization requirements. The above factors reduce the memory access rate; for instance, under the ideal assumptions of a 100\% cache hit rate, 100\% DRAM row-buffer hit rate, and perfect interleaving, fully utilizing our evaluated system with 204.8 GB/s Last Level Cache (LLC) bandwidth and 51.2 GB/s DDR4 memory bandwidth requires one LLC request per cycle and one DRAM request every four cycles. \maa{} addresses these issues by leveraging dedicated compute units for address computation, generating hundreds of outstanding transactions to mask latency, and using shared scratchpad memory for low-latency RMW operations.
}
\vspace{-2mm}

\section{\maa{} Architecture}~\label{sec:architecture}

\vspace{-2mm}

\aptLtoX[graphic=no,type=html]{\begin{table}[t]
    \footnotesize
    \centering
    \captionof{table}{Common Data Access Patterns of Irregular Applications}
     \vspace{-3mm}
    \setlength{\tabcolsep}{2pt}
        \begin{tabular}{|c|c|c|c|c|}
            \hline 
            \textbf{Benchmark} & \textbf{Kernel} & \textbf{Access} & \textbf{Condition/Address Calculation} & \textbf{Loop} \\
            \hline \hline
            \multirow{2}{*}{NAS~\cite{nas}} & IS & RMW \texttt{A[B[i]]} & -- & \texttt{i = F to G} \\
            \cline{2-5}
             & CG & LD \texttt{A[B[j]]} & -- & \texttt{j = H[i] to H[i+1]} \\
            \hline
            \multirow{3}{*}{GAP~\cite{gap}} & BFS & ST \texttt{A[B[j]]} & \texttt{if (D[E[j]] < F)} & \texttt{j = H[K[i]] to H[K[i]+1]} \\
            \cline{2-5}
             & BC & RMW \texttt{A[B[j]]} & \texttt{if (D[E[j]] == F)} & \texttt{j = H[K[i]] to H[K[i]+1]} \\
            \cline{2-5}
             & PR & RMW \texttt{A[B[j]]} & -- & \texttt{j = H[i] to H[i+1]} \\
            \hline
            \multirow{2}{*}{Hash-Join~\cite{hashjoin}} & PRH & ST \texttt{A[B[f(C[i])]]} & \texttt{f(C[i]) = (C[i] \& F) \mbox{>}\mbox{>} G} & \texttt{i = F to G} \\
            \cline{2-5}
             & PRO & ST \texttt{A[B[f(C[i])]]} & \texttt{f(C[i]) = (C[i] \& F) \mbox{>}\mbox{>} G} & \texttt{i = F to G} \\
            \hline
            \multirow{4}{*}{UME~\cite{ume}} & GZZ & RMW \texttt{A[B[[i]]]} & \texttt{if (D[i] >= F)} & \texttt{i = F to G} \\
            \cline{2-5}
            & GZZI & LD \texttt{A[B[C[j]]]} & \texttt{if (D[j] >= F)} & \texttt{j = H[K[i]] to H[K[i]+1]} \\
            \cline{2-5}
            & GZP & RMW \texttt{A[B[i]]} & \texttt{if (D[i] >= F)} & \texttt{i = F to G} \\
            \cline{2-5}
            & GZPI & LD \texttt{A[B[C[j]]]} & \texttt{if (D[j] >= F)} & \texttt{j = H[K[i]] to H[K[i]+1]} \\
            \hline
            Spatter~\cite{spatter} & XRAGE & ST \texttt{A[B[i]]} & -- & \texttt{i = F to G} \\
            \hline
            \multicolumn{5}{l}{*We show only one pattern for each kernel because of space limitation.}
        \end{tabular}
    \label{tab:pattern}
\end{table}
\begin{table}
\footnotesize
    \centering
    \captionof{table}{\maa{} Instruction Set}
     \vspace{-3mm}
    \renewcommand{\arraystretch}{1.1}
    \label{tab:isa}
    \begin{tabular}{|c|c|l|}
        \hline
        \textbf{Type} & \textbf{Opcode} & \textbf{Operands} \\
        \hline
        \hline
        \multirow{3}{*}{\shortstack{Indirect \\ Access}} & \texttt{ILD} & \texttt{DTYPE BASE TD TS1 TC} \\
        \cline{2-3} 
        & \texttt{IST} & \texttt{DTYPE BASE TS1 TS2 TC} \\
        \cline{2-3} 
        & \texttt{IRMW} & \texttt{DTYPE BASE OP TS1 TS2 TC} \\
        \hline
        \hline
        Stream & \texttt{SLD} & \texttt{DTYPE BASE TD RS1 RS2 RS3 TC} \\ 
        \cline{2-3}
        Access & \texttt{SST} & \texttt{DTYPE BASE TS RS1 RS2 RS3 TC} \\
        \hline
        \hline
        \multirow{2}{*}{ALU} & \texttt{ALUV} & \texttt{DTYPE OP TD TS1 TS2 TC} \\
        \cline{2-3}
        & \texttt{ALUS} & \texttt{DTYPE OP TD TS RS TC} \\
        \hline
        \hline
        Range & \multirow{2}{*}{\texttt{RNG}} & \multirow{2}{*}{\texttt{TD1 TD2 TS1 TS2 RS1 TC}} \\
        Loop & & \\
        \hline
    \end{tabular}
\end{table}}{\begin{table*}[t]
\begin{minipage}[t]{0.57\textwidth}
    \footnotesize
    \centering
    \captionof{table}{Common Data Access Patterns of Irregular Applications}
     \vspace{-3mm}
    \setlength{\tabcolsep}{2pt}
        \begin{tabular}{|c|c|c|c|c|}
            \hline 
            \textbf{Benchmark} & \textbf{Kernel} & \textbf{Access} & \textbf{Condition/Address Calculation} & \textbf{Loop} \\
            \hline \hline
            \multirow{2}{*}{NAS~\cite{nas}} & IS & RMW \texttt{A[B[i]]} & -- & \texttt{i = F to G} \\
            \cline{2-5}
             & CG & LD \texttt{A[B[j]]} & -- & \texttt{j = H[i] to H[i+1]} \\
            \hline
            \multirow{3}{*}{GAP~\cite{gap}} & BFS & ST \texttt{A[B[j]]} & \texttt{if (D[E[j]] < F)} & \texttt{j = H[K[i]] to H[K[i]+1]} \\
            \cline{2-5}
             & BC & RMW \texttt{A[B[j]]} & \texttt{if (D[E[j]] == F)} & \texttt{j = H[K[i]] to H[K[i]+1]} \\
            \cline{2-5}
             & PR & RMW \texttt{A[B[j]]} & -- & \texttt{j = H[i] to H[i+1]} \\
            \hline
            \multirow{2}{*}{Hash-Join~\cite{hashjoin}} & PRH & ST \texttt{A[B[f(C[i])]]} & \texttt{f(C[i]) = (C[i] \& F) \mbox{>}\mbox{>} G} & \texttt{i = F to G} \\
            \cline{2-5}
             & PRO & ST \texttt{A[B[f(C[i])]]} & \texttt{f(C[i]) = (C[i] \& F) \mbox{>}\mbox{>} G} & \texttt{i = F to G} \\
            \hline
            \multirow{4}{*}{UME~\cite{ume}} & GZZ & RMW \texttt{A[B[[i]]]} & \texttt{if (D[i] >= F)} & \texttt{i = F to G} \\
            \cline{2-5}
            & GZZI & LD \texttt{A[B[C[j]]]} & \texttt{if (D[j] >= F)} & \texttt{j = H[K[i]] to H[K[i]+1]} \\
            \cline{2-5}
            & GZP & RMW \texttt{A[B[i]]} & \texttt{if (D[i] >= F)} & \texttt{i = F to G} \\
            \cline{2-5}
            & GZPI & LD \texttt{A[B[C[j]]]} & \texttt{if (D[j] >= F)} & \texttt{j = H[K[i]] to H[K[i]+1]} \\
            \hline
            Spatter~\cite{spatter} & XRAGE & ST \texttt{A[B[i]]} & -- & \texttt{i = F to G} \\
            \hline
            \multicolumn{5}{l}{*We show only one pattern for each kernel because of space limitation.}
        \end{tabular}
    \label{tab:pattern}
\end{minipage}%
\begin{minipage}[t]{0.50\textwidth}
\footnotesize
    \centering
    \captionof{table}{\maa{} Instruction Set}
     \vspace{-3mm}
    \renewcommand{\arraystretch}{1.1}
    \label{tab:isa}
    \begin{tabular}{|c|c|l|}
        \hline
        \textbf{Type} & \textbf{Opcode} & \textbf{Operands} \\
        \hline
        \hline
        \multirow{3}{*}{\shortstack{Indirect \\ Access}} & \texttt{ILD} & \texttt{DTYPE BASE TD TS1 TC} \\
        \cline{2-3} 
        & \texttt{IST} & \texttt{DTYPE BASE TS1 TS2 TC} \\
        \cline{2-3} 
        & \texttt{IRMW} & \texttt{DTYPE BASE OP TS1 TS2 TC} \\
        \hline
        \hline
        Stream & \texttt{SLD} & \texttt{DTYPE BASE TD RS1 RS2 RS3 TC} \\ 
        \cline{2-3}
        Access & \texttt{SST} & \texttt{DTYPE BASE TS RS1 RS2 RS3 TC} \\
        \hline
        \hline
        \multirow{2}{*}{ALU} & \texttt{ALUV} & \texttt{DTYPE OP TD TS1 TS2 TC} \\
        \cline{2-3}
        & \texttt{ALUS} & \texttt{DTYPE OP TD TS RS TC} \\
        \hline
        \hline
        Range & \multirow{2}{*}{\texttt{RNG}} & \multirow{2}{*}{\texttt{TD1 TD2 TS1 TS2 RS1 TC}} \\
        Loop & & \\
        \hline
    \end{tabular}
\end{minipage}
\vspace{-4mm}
\end{table*}}

\begin{figure}[t]
    \centering
    \includegraphics[width=0.95\linewidth]{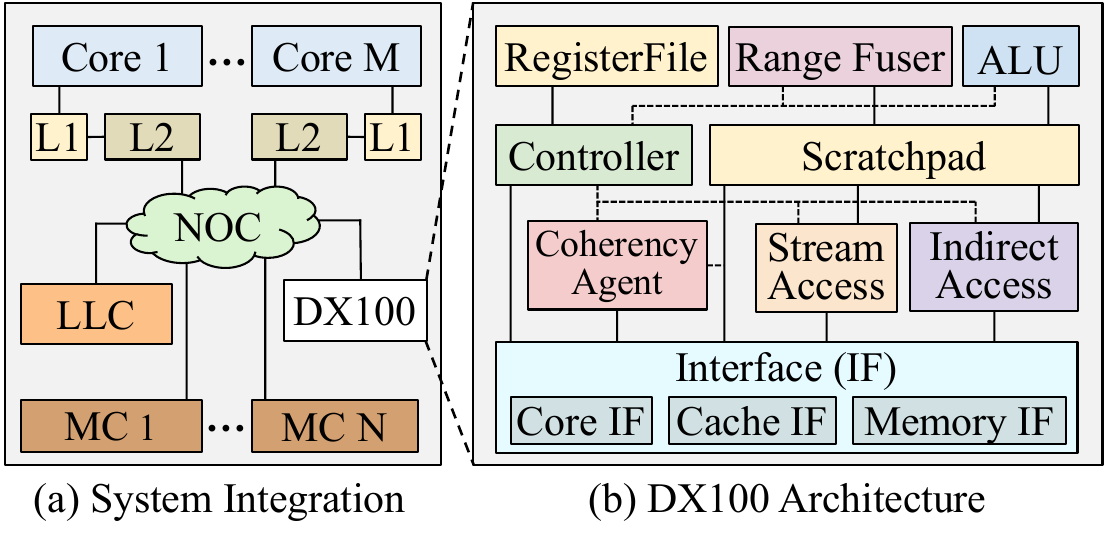}
     \vspace{-4mm}
    \caption{(a) \maa{} is integrated as a shared memory-mapped accelerator to the system. (b) \maa{} architecture.}
    \Description[\maa{} architecture overview.]{(a) shows that \maa{} is integrated as a shared memory-mapped accelerator to the system. (b) depicts \maa{} architecture.}
    \vspace{-2mm}
    \label{fig:architecture_overview}
\end{figure}

This section delves into the details of \maa{} microarchitecture. We provide an overview first, before describing individual functional units and system interfaces for the accelerator. \maa{} executes hoisted bulk memory accesses at a granularity of \textbf{tile} (\textit{e.g.,} 16K 4B words). The CPU cores perform all remaining computations. Figure~\ref{fig:architecture_overview} (a) illustrates the placement of the \maa{} as a shared accelerator within the processor. \maa{} is integrated as a modular memory-mapped component within the coherent fabric connected via the network-on-chip to system components, requiring minimal adjustments to the core, memory controller, or instruction-set architecture. \maa{} is designed as an independent component for flexible integration with mesh~\cite{skylake_hotchips,skylake_isscc} or ring~\cite{zen4_hotchips} interconnects.

\maa{} is shared among a group of cores within the processor. While we focus on a single \maa{} instance throughout the paper, we explore the scalability of multiple instances in Section~\ref{subsec:scalability}.
Compared to previous fetcher units that use private co-processors, the shared design offers several key benefits:
\textit{First}, the area overhead of a shared accelerator is amortized across multiple cores. A memory-mapped accelerator requires minimal modifications to the core, as opposed to the co-processors, which introduce new instructions to the ISA or modify the LSQ.
\textit{Second}, a shared accelerator can reorder memory accesses across threads from different cores in a shared-memory environment (\textit{e.g.,} Pthread or OpenMP), further improving row buffer hit rates and enabling higher coalescing factors.
\textit{Third}, \maa{} does not require fine-grain atomic accesses for store and RMW operations as it maintains exclusive access to the indirect memory regions. 
Considering these advantages, we designed \maa{} as a shared accelerator.
The primary drawback of this approach is slower communication between cores and the accelerator.
To mitigate this, we make the core's \textit{data} accesses to \maa{} (\textit{e.g.}, for loading gathered data) cacheable, enabling stride prefetchers to effectively reduce the data access latency (Section~\ref{subsec:interface_CA}).

Figure~\ref{fig:architecture_overview} (b) provides a high-level overview of the \maa{} architecture. \textit{Scratchpad} contains multiple tiles that are the source and destination of \maa{} instructions, facilitating communication between functional units and between \maa{} and cores. All functional units within \maa{} operate in \texttt{TILE} granularity (\textit{e.g.}, 16K 4B words). \textit{Stream Access} performs streaming loads and stores (\textit{e.g.,} \texttt{B[i]}) from the LLC to the Scratchpad. \textit{Indirect Access} reads index values from the Scratchpad and performs indirect load, store, or RMW (\textit{e.g.,} \texttt{A[B[i]]}) from LLC or memory. \textit{Range Fuser} combines multiple small range loops with few iterations into a single larger loop for efficient bulk access with a large tile size. \textit{ALU Unit} performs condition calculations and index value transformations.
\textit{Controller} manages the dispatch, issue, and retirement of \maa{} instructions.
\textit{Coherency Agent} tracks cached scratchpad data to ensure coherence.
\textit{Interface} manages \maa{} accesses to the cache and memory, as well as core accesses to \maa{}.
Finally, \textit{Register File} holds scalar values required for loops and ALU operations.

\subsection{Common Patterns and ISA}~\label{subsec:common_patterns}
Table~\ref{tab:pattern} demonstrates access, condition, address calculation, and loop patterns in evaluated irregular workloads with the following key takeaways:
\textit{First}, in addition to indirect loads (\texttt{C = A[B[i]]}), store (\texttt{A[B[i]] = C}) and read-modify-write (RMW) operations (\texttt{A[B[i]] += C}) are frequently used to update values at indirect addresses.
\textit{Second}, supporting stores and RMWs is challenging because they are often conditioned.
Unlike loads, where ignoring conditions may only result in performance degradation by fetching extra data, neglecting conditions for store and RMW compromises the correctness.
\textit{Third}, besides single loops (\texttt{i = F to G}), range loops are commonly employed in two forms:
direct range (\texttt{j = H[i] to H[i+1]}) where the loop induction variable \texttt{j} increments sequentially, or indirect range (\texttt{j = H[K[i]] to H[K[i]+1]}) where the ranges themselves depend on indirect accesses.
\textit{Fourth}, indirect range loops usually cover only a few iterations.
A common example is graph workloads with frontier queues, where a range loop iterates over a few neighbors of the current frontier nodes. \textit{Finally}, multiple levels of indirection (\texttt{A[B[C[i]]]}) are common. 

\maa{} ISA seamlessly supports these patterns with eight instructions as shown in Table~\ref{tab:isa}. The ISA
(1) \textit{Decouples} streaming accesses from indirect accesses. Streaming accesses use scratchpad tiles (\texttt{TD}) as the destination.
Indirect accesses use scratchpad tiles as index source (\texttt{TS}).
(2) Supports different access types (\texttt{LD, ST, RMW}).
(3) Allows various \texttt{ALU} operations (\texttt{OP} operand) for conditional memory accesses using the condition tile (\texttt{TC}).
(4) Enables fusing range loops (\texttt{RNG}) for bulk accesses.
(5) Supports various data types (\texttt{DTYPE}), such as \textit{u32}, \textit{i32}, \textit{f32}, \textit{u64}, \textit{i64}, and \textit{f64}, and different operations (\texttt{OP}), such as \textit{ADD}, \textit{SUB}, \textit{MUL}, \textit{MIN}, \textit{MAX}, \textit{AND}, \textit{OR}, \textit{XOR}, \textit{SHR}, \textit{SHL}, \textit{LT}, \textit{LE}, \textit{GT}, \textit{GE}, and \textit{EQ} required for a general-purpose usage.
Note, \maa{} only supports a subset of associative and commutative operations, such as \textit{ADD}, \textit{MAX}, and \textit{MIN} for the \textit{IRMW} instructions as they reorder the operations.
These features, using \textit{scratchpad as an intermediate storage}, enable offloading conditional loops, multiple levels of indirection, and complex address calculations.
\vspace{-2mm}
\subsection{Indirect Access Unit}
\begin{figure*}[t]
    \centering
    \includegraphics[width=0.98\textwidth, trim=0mm 0cm 1mm 0cm, clip]{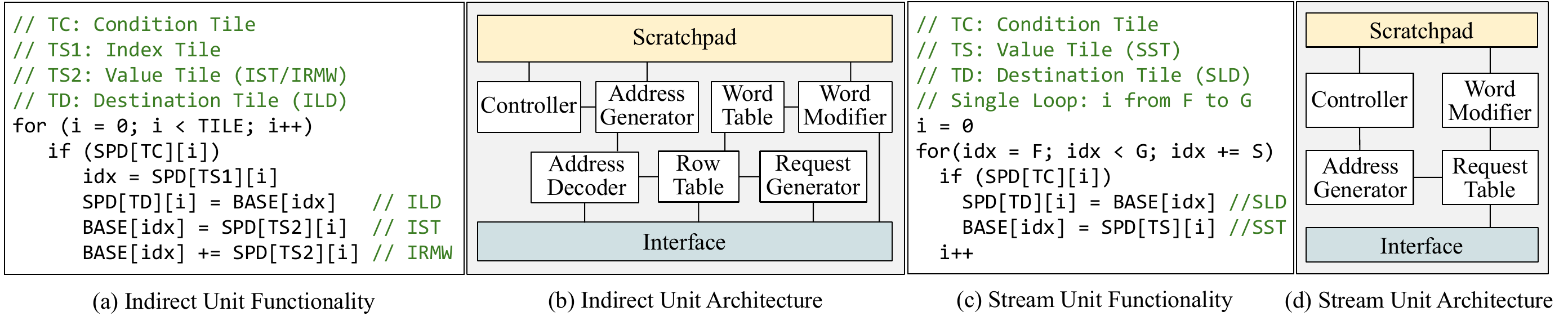}
     \vspace{-5mm}
    \caption{Stream and Indirect Access unit functionality and 
    architecture.}
    \Description[Stream and Indirect Access unit functionality and architecture.]{(a) shows that the Indirect Access unit performs indirect load, store, and read-modify-write. (b) depicts the architecture of the Indirect Access unit. (c) shows that the Stream Access unit performs streaming loads and stores. (d) depicts the architecture of the Stream Access unit.}
    \label{fig:stream_architecture}
     \vspace{-2mm}
\end{figure*}

\textbf{Functionality:} Figure~\ref{fig:stream_architecture} (a) shows the functionality of the Indirect Access unit, which performs load, store, and RMW operations on indirect memory locations. The index tile (\texttt{TS1}) has been fetched and exists in the scratchpad. \texttt{TS1} can originate from streaming access for single-level indirection (\texttt{A[B[i]]}), indirect access for multi-level indirection (\texttt{A[B[C[i]]]}), or the ALU unit for complex address calculations (\texttt{A[f(C[i])]}).

\noindent\textbf{Memory Bandwidth Enhancements:} Figure~\ref{fig:stream_architecture} (b) details the architecture of this unit, which is designed to optimize DRAM bandwidth utilization.
The Indirect Access unit employs a Row Table to \textit{reorder} indirect addresses by storing addresses belonging to the same DRAM row together, and issuing them at the same time. 
This approach ensures that all accesses to columns within the same DRAM row are completed consecutively, increasing the row buffer hit rate.
Additionally, a Word Table organizes target words within each column into a linked list.
This component enables the unit to access unique columns, effectively \textit{coalescing} redundant accesses within a tile.
Lastly, Request Generator \textit{interleaves} accesses across different DRAM channels and bank groups, maximizing channel bandwidth and enabling bank-group interleaving.

\begin{figure}[t]
    \centering
    \includegraphics[width=0.90\linewidth]{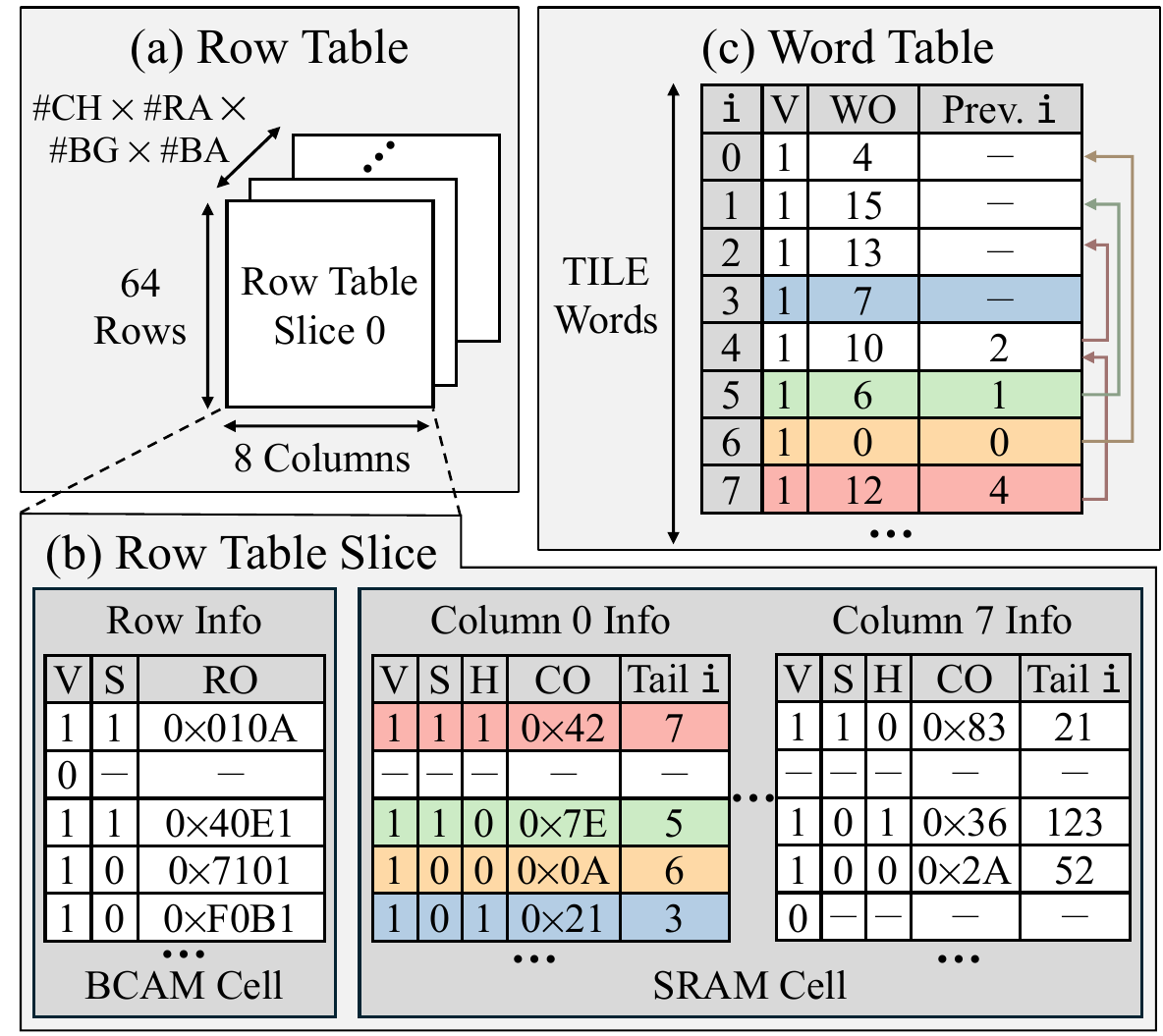}
    \vspace{-3mm}
    \caption{(a) Row Table, (b) Row Table Slice, and (c) Word Table architecture of Indirect Access unit.}
    \Description[Row Table and Word Table architecture.]{(a) shows that the Row Table contains a slice for each DRAM bank. (b) shows the architecture of a Row Table slice with a BCAM cell and an SRAM cell to maintain the row and column addresses. (c) shows the architecture of the Word Table, which contains the word ID of each element within a tile and creates a linked list of the words within a cache line.}
    \vspace{-4mm}
    \label{fig:row_word_table}
\end{figure}

\noindent\textbf{Row Table Architecture.}
Figure~\ref{fig:row_word_table} (a) shows the Row Table, which consists of multiple slices corresponding to the DRAM banks.
This design enables high-throughput lookups for row and column addresses.
Figure~\ref{fig:row_word_table} (b) shows the slice architecture, which stores row and column information for outstanding requests to a specific DRAM bank. Each slice includes a Binary Content Addressable Memory (BCAM) and an SRAM cell.
BCAM offers fully associative lookup access to the information of 64 target DRAM rows.
It stores a valid bit (\texttt{V}), a sent bit (\texttt{S}) indicating whether all requests to the row have been issued, and the row address (\texttt{RO}).
SRAM cell holds the information of up to 8 DRAM columns per row, such as a valid bit (\texttt{V}), a sent bit (\texttt{S}), a cache hit bit (\texttt{H}) indicating if the column is present in the caches, the column address (\texttt{CO}), and a pointer to the Word Table (\texttt{Tail i}) targeting words within the column.

\noindent\textbf{Word Table Architecture.}
Figure~\ref{fig:row_word_table} (c) depicts the Word Table, which holds the word information in target DRAM columns using a linked-list structure.
For each word in the tile corresponding to iteration number \texttt{i}, Word Table stores a valid bit (\texttt{V}), the word's offset within the column (\texttt{WO}), and the previous iteration (\texttt{Previous i}) accessing the same column, enabling the linked-list organization.

The Indirect Access unit operates in three stages:
\textbf{Operation Stage 1 - Fill:}
Controller reads the condition values (\texttt{SPD[TC][i]}) from the Scratchpad.
If condition holds, it triggers Address Generator with the iteration number (\texttt{i}).
Address Generator fetches the index value (\texttt{idx=SPD[TS1][i]}) from the Scratchpad and calculates the virtual word address (\texttt{BASE[idx]}).
Address Decoder translates the virtual address to the physical address using Interface.
Then, it maps the physical address to DRAM coordinates (\texttt{CH}, \texttt{RA}, \texttt{BG}, \texttt{BA}, \texttt{RO}, and \texttt{CO}) and word offset (\texttt{WO}), which are inserted into Row and Word Tables in 4 steps: (a) The Row Table slice is determined based on the \texttt{CH}, \texttt{RA}, \texttt{BG}, and \texttt{BA} target addresses.
(b) The BCAM cell of the identified slice looks for a valid and unsent entry with the target row address \texttt{RO}.
Once the row is selected, the SRAM cell finds a valid and unsent entry with the target column address \texttt{CO}, captures its \texttt{Tail i} field, and updates it to the iteration number of the current target element.
If no such entry exists in the BCAM or SRAM cells, the Row Table allocates a new entry.
When a DRAM column is accessed for the first time, the Row Table queries the Interface to determine the cache hit status (\texttt{H} bit, see Section~\ref{subsec:interface_CA}).
(c) The new word offset \texttt{WO} is inserted into the \texttt{i}$^{th}$ entry of the Word Table.
The \texttt{Previous i} field of this entry is set with \texttt{Tail i} from Row Table, linking the current word to the prior words in the same column.

\noindent\textbf{Operation Stage 2 - Request:}
Once all words are inserted for a row or the Row Table reaches capacity, the Indirect Access unit initiates the process of generating and issuing memory accesses to the Interface.
For that, each Row Table slice sequentially scans its BCAM and SRAM cells to identify the valid and unsent \texttt{RO} and \texttt{CO} addresses, as well as the cache hit bit (\texttt{H}).
The Request Generator arbitrates among Row Table slices in a predetermined order, interleaving slices corresponding to different DRAM channels and bank groups for optimized bandwidth utilization.
Using \texttt{RO} and \texttt{CO} addresses, and the slice's bank coordinates, the Request Generator calculates the target memory address.
This information, along with the \texttt{H} bit, is then sent to the Interface to load the required data.

\noindent\textbf{Operation Stage 3 - Response:}
When a cache line is received from the Interface, the Address Decoder maps its address to DRAM coordinates.
The Row Table looks for valid and sent columns in its corresponding slice to identify the tail of the word linked-list (\texttt{Tail i} field).
The Word Table traverses this linked-list to retrieve the word offsets (\texttt{WO}) and iteration numbers (\texttt{i}).
Word Modifier then processes the data based on the instruction type:
\textit{Indirect Load} (\texttt{ILD}): It extracts the required words from the response data and writes them to the destination Scratchpad tile (\texttt{TD}).
\textit{Indirect Store} (\texttt{IST}): It reads the new words from the source tile (\texttt{TS2}), and inserts them into the response data.
\textit{Indirect RMW} (\texttt{IRMW}): It reads new words from the source tile, extracts words from the response data, performs arithmetic operations, and inserts the modified words back into the response data.
For store and RMW operations, modified data is sent back to the Interface to complete the write operation.

\subsection{Stream Access Unit}
Figure~\ref{fig:stream_architecture} (c) shows the functionality of the Stream Access Unit.
The streaming load (\texttt{SLD}) instruction loads a tile of data elements from sequential addresses in a single loop and writes them to the scratchpad.
The streaming store (\texttt{SST}) instruction stores a tile of values from the scratchpad to sequential memory addresses.
Figure~\ref{fig:stream_architecture} (d) demonstrates the architecture of this unit.
Controller generates loop iteration values (\texttt{idx} and \texttt{i}).
In conditioned instructions, Controller retrieves condition values from the scratchpad and skips iterations where the condition is not met.
Address Generator computes the streaming virtual cache line addresses and word offsets (\texttt{wid}) within each cache line.
Request Table functions similarly to MSHR by tracking outstanding memory addresses and their \texttt{wid} and \texttt{i} values.
It performs the address translation using the Interface, generates load requests, and injects them back into the Interface.
Upon receiving responses, Request Table retrieves the associated \texttt{wid} and \texttt{i} values.
For \texttt{SLD}, Word Modifier extracts specific words from the cache line and writes them to the scratchpad.
For \texttt{SST}, Word Modifier reads the new values from the scratchpad, inserts them to the cache line, and issues a store request to the Interface.

Finally, note that the cores can efficiently handle streaming accesses (\texttt{B[i]}) and some address calculation operations.
Performing these tasks on the core and then transferring the data to \maa{} would simplify the accelerator design, allowing it to focus specifically on indirect accesses with lower memory bandwidth utilization.
However, this approach would increase the data transfer between the core and \maa{}, making the \maa{} interface a bottleneck.
As a result, the accelerator would frequently stall while waiting for the required data for indirect access.
This reduces the rate of \maa{} memory requests and the memory bandwidth utilization. Therefore, we decided to support all these patterns locally within \maa{}.

\subsection{Range Fuser and ALU Units}

\begin{figure}[h]
    \centering
    \vspace{-3mm}
    \includegraphics[width=0.8\linewidth]{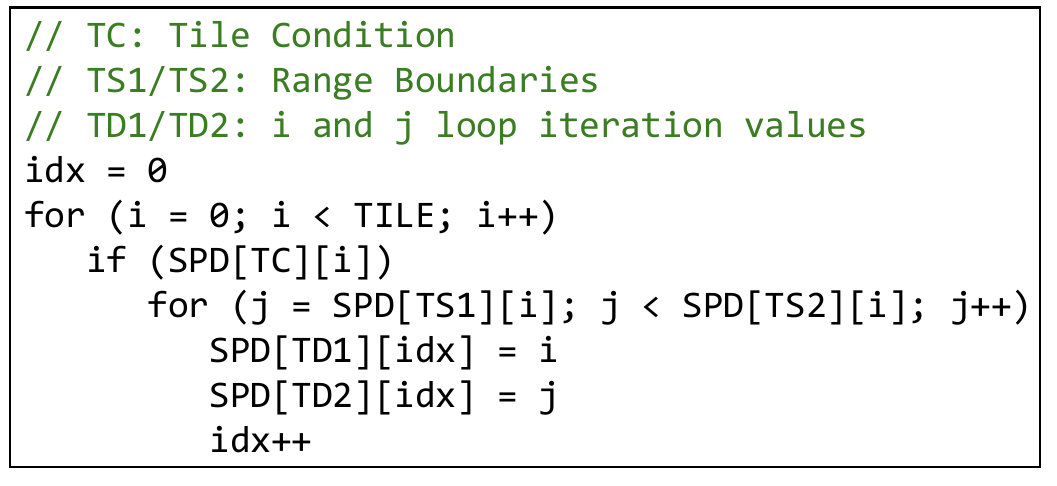}
    \vspace{-4mm}
    \caption{Range Fuser Functionality.}
    \Description[Range Fuser Functionality.]{Range Fuser combines multiple range loops with a few iterations for bulk data accesses. It extracts the i and j loop induction variables from the inner loop to two tiles. These tiles can be further used by the indirect access unit for indirect data access.}
    \vspace{-3mm}
    \label{fig:range}
\end{figure}

Section~\ref{subsec:common_patterns} showed that range loops like \texttt{j = H[K[i]] to H[K[i]+1]} involve few iterations, which is inefficient for offloading bulk accesses to \maa{}.
To address this limitation, we provision a \textbf{Range Fuser} unit, which combines multiple small-range loops into a single larger range.
Figure~\ref{fig:range} illustrates the functionality of the Range Fuser unit.
It processes two input tiles representing the minimum and maximum range boundaries (\texttt{H[K[i]]} and \texttt{H[K[i]+1]}).
These boundaries can be loaded either by the Stream Access unit for direct range loops or by the Indirect Access unit for indirect range loops.
The Range Fuser generates two output tiles that contain the induction variable values of the outer single loop (\texttt{i}) and the inner range loop (\texttt{j}).
These tiles can then be passed to the Indirect Access unit to facilitate further memory accesses, such as \texttt{A[B[j]]}.

The \textbf{ALU} unit executes arithmetic, bitwise, and comparison operations required for condition evaluation and address calculations (refer to Table~\ref{tab:pattern}).
The ALU operands can be two tiles (\texttt{D[i] > E[i]}) or a tile and a scalar value from the register file (\texttt{C[i] \& F}).
The output condition tiles can be utilized by all functional units for subsequent operations, and address calculation results can be used as indices for indirect memory accesses.

\vspace{-2mm}
\subsection{Controller, Register File, and Scratchpad}~\label{subsec:controller_RF_SPD}
\textbf{Controller} receives \maa{} instructions from the cores and schedules them through multiple stages: dispatch, issue, execute, and retire.
Each \maa{} instruction is 192b wide and is transmitted via three 64b memory-mapped stores.
To support out-of-order execution, Controller employs a scoreboard.
To avoid WAW and RAW hazards without register renaming, Controller prevents the dispatch of instructions targeting destination tiles that are already in use by a valid instruction in the scoreboard.

\textbf{Register File} contains scalar values, which are used for single-loop boundaries, loop strides, and ALU operations.
\textbf{Scratchpad} stores multiple tiles, each containing \texttt{TILE} elements.
For each tile, Scratchpad maintains a size and a ready bit.
When an instruction is dispatched, ready bits corresponding to all source and destination tiles are set to 0, indicating that the tiles have not finished the execution yet.
Once the instruction retires, these bits are reset to 1.
This bit allows the synchronization between cores with \maa{}.

In addition to ready bits, Scratchpad maintains a finish bit for all elements.
When an instruction is issued to a functional unit, the finish bits of all destination tile elements are set to 0.
As the elements are computed and written back to the Scratchpad, their finish bits are set to 1.
This design enables fine-grained execution coordination between functional units.
For example, as soon as an element of \texttt{B[i]} is loaded into the Scratchpad by the Stream Access unit, the Indirect Access unit can immediately access it and populate the Row Table and Word Table for memory operations like \texttt{A[B[i]]}.
This overlapping of operations hides the fill latency of the Indirect Access unit behind the load of the index array.

\vspace{-2mm}
\subsection{Interface and Coherency Agent}~\label{subsec:interface_CA}
Interface manages \maa{} memory accesses using the \textbf{Cache Interface} and \textbf{DRAM Interface}.
Two implementation options were considered:
(a) \textit{Injecting accesses into the LLC:}
This ensures memory accesses follow the coherency protocol, benefiting from lower latency and higher LLC bandwidth if they hit.
Misses are forwarded to memory controllers by the LLC MSHRs.
However, LLC and NoC reorder packets, which can negate the row buffer hit rate improvements achieved by the Indirect Access unit, particularly in mesh-based fabrics like Intel~\cite{skylake_hotchips,skylake_isscc}.
(b) \textit{Directly accessing memory:}
This preserves the order of indirect accesses and avoids LLC MSHR limitations on memory-level parallelism.
However, this approach requires the interface to handle coherency.
Our design leverages the strengths of both methods.
Streaming accesses, which exhibit high spatial and temporal locality, are directed to the LLC using the Cache Interface.
Indirect accesses, prone to frequent LLC misses, can bypass the LLC and access memory directly.
To maintain coherency, Interface snoops the coherency directories during the fill stage to check if the required cache line is valid, storing this information in the \texttt{H} bit of the Row Table.
In the request stage, the interface routes the access to the LLC using the Cache Interface if the H bit is set; otherwise, it issues the access directly to memory controllers using the DRAM Interface.
This approach ensures correctness, as \maa{} maintains exclusive write access to the indirect arrays within the ROI (Section~\ref{subsec:compiler} -- Legality).
Thus, no cores can modify the cache lines between snooping and issuing the request.

\setlength{\intextsep}{0pt} 
\setlength{\columnsep}{10pt} 
\begin{wrapfigure}{R}{0.35\linewidth}
    \captionsetup{skip=2pt}
    \centering
    \includegraphics[width=\linewidth, trim=0mm 0mm 5mm 0cm, clip]{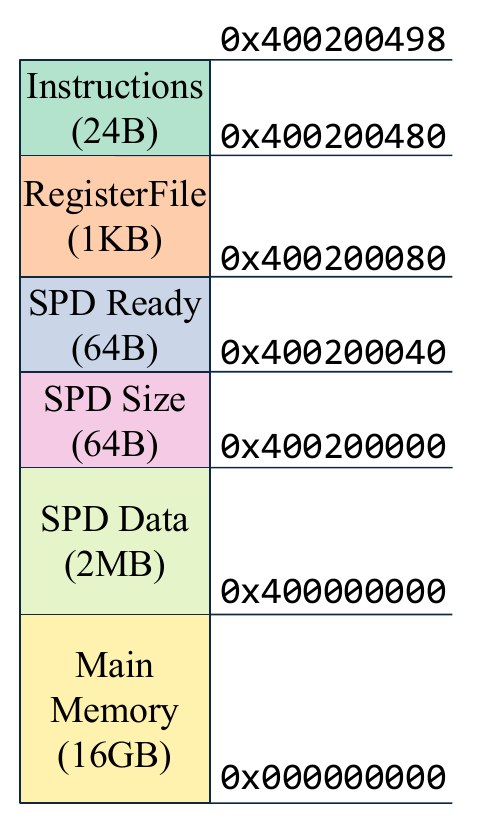}
    \centering
    \caption{\maa{} \\ Memory Regions.}
    \label{fig:memory_region}
    \Description[\maa{} memory-mapped regions.]{\maa{} exposes several memory-mapped regions: 2MB for scratchpad data, 64B for tile size, 64B for tile ready bits, 1KB for register file data, and 24B for instruction reception.}
\end{wrapfigure}
Interface also contains the \textbf{Core Interface} to handle core access to \maa{} for sending instructions and reading/writing to the Register File (RF) or Scratchpad (SPD).
Figure~\ref{fig:memory_region} shows how these memory regions are mapped to the host memory address space, using the system configuration in Table~\ref{tab:config}.
All \maa{} memory regions are uncacheable, except for scratchpad data.
This region is accessed in a streaming, non-temporal manner. Hence, cache stride prefetchers efficiently prefetch scratchpad data, reducing core access latency.
To handle cacheable scratchpad accesses, \maa{} includes a \textbf{Coherency Agent} that tracks scratchpad cache line status using a single valid bit (\texttt{V}) per cache line.
When cores read from the scratchpad, the \texttt{V} bit is set. When dispatching instructions, \maa{} Controller triggers the coherency agent to invalidate all scratchpad cache lines associated with the source and destination tiles of the instruction.

\textbf{Address Translation.}
We map the virtual page addresses of \maa{} memory regions to the same physical addresses as a general approach for memory-mapped IO communication.
We assume the stream and indirect memory regions are mapped through huge pages, which is a common solution for reducing the address translation overhead of applications with large datasets~\cite{huge_page, huge_page2, huge_page3, huge_page4}.
Using huge pages, we provision a small 256-entry TLB to keep the Page Table Entries (PTE) of the required memory regions.
We provide certain \maa{} APIs (Section~\ref{subsec:programmint_API}) to transfer the required PTEs to \maa{}'s TLB once for the whole application lifetime.










\vspace{-2mm}
\section{Programming Model and Compiler}
\label{sec:programming}
We provide two methods to use \maa{} in software:
a library of APIs that require manual insertion by the programmer and an automatic compiler system.
Due to compiler limitations (memory dependence analysis and code pattern detection), the manual programming method can serve as a fallback and ensure higher code coverage.

\vspace{-2mm}
\subsection{Manual Programming API}~\label{subsec:programmint_API}
Section~\ref{subsec:interface_CA} shows that the instructions are transmitted to \maa{} using three 64-bit atomic stores targeting specific \maa{} memory regions.
To facilitate this process, we developed a library of \maa{} APIs that perform instruction encoding, memory-mapped accesses, page table entry transfers, and tile/register allocation for various supported data types.
Additionally, the library provides a \texttt{wait} instruction for core-\maa{} synchronization. When invoked, the core will poll the corresponding Scratchpad's ready bit until it is set (refer to Section~\ref{subsec:controller_RF_SPD}).
These APIs enable users or compiler infrastructure to seamlessly transform the legacy code for \maa{} integration.
Figures~\ref{fig:compiler_trans_eg} (a) and (d) show a simple gather code and its offloaded \maa{} version using these APIs (noted by \textbf{\texttt{\maa{}}}).

\textbf{Limitations.}
\maa{}'s support for data access acceleration is limited by the following factors:
\textit{Semantic limitation:} Conventional pointer-chasing and linked-list traversal operations involve pointer dereferencing with random base addresses (\textit{e.g.}, \texttt{*next[i]}). \maa{} does not support this pattern, as our workload analysis did not reveal its usage—likely due to memory fragmentation concerns. For instance, the bucket-chaining join~\cite{bucketchaining_pro} algorithm of Hash-Join benchmark~\cite{hashjoin} uses array-based indirection (\textit{e.g.}, \texttt{nodes[next\_idx[i]]}) for linked-list traversal. \maa{} accelerates this pattern by processing bulk linked-list traversal operations across many tuples.
\textit{Parallelism limitation:} \maa{} requires \emph{bulk} accesses for offloading and reordering. In graph workloads~\cite{gap}, the parallelism is determined by the number of nodes in the frontier list. When an iteration involves a few nodes, \maa{} acceleration is not beneficial due to the tile under-utilization. In these iterations, we revert to the baseline C code. Across all evaluated graph algorithms, \maa{} acceleration covers more than 99\% of nodes, while the remaining nodes are processed using the non-accelerated code.

\subsection{Compiler}~\label{subsec:compiler}
Identifying indirect access patterns and rewriting the whole program using provided APIs could be complicated. Therefore, we propose an MLIR-based compiler to alleviate this problem.

\begin{figure}[ht]
    \centering
    \includegraphics[width=0.85\linewidth]{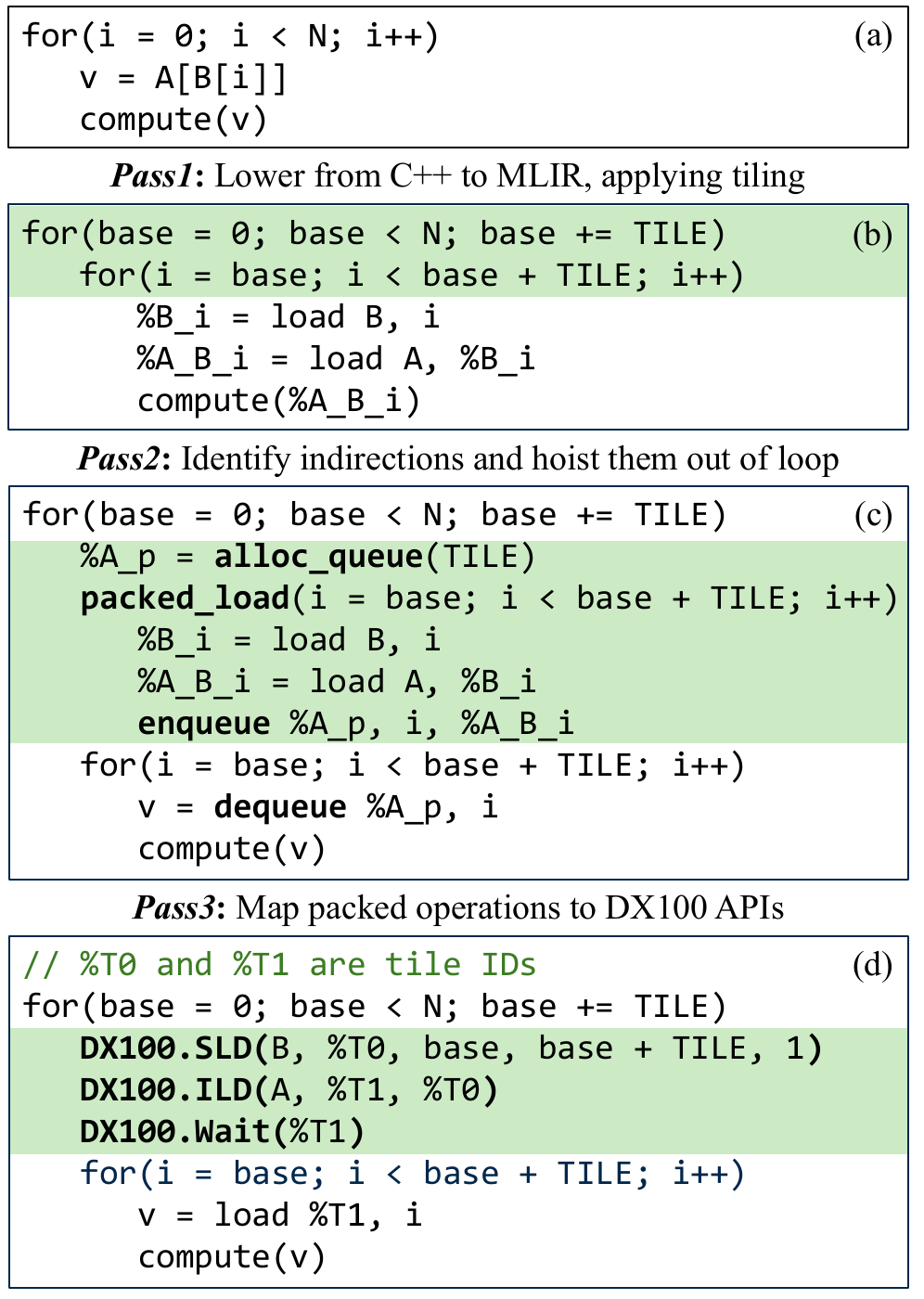}
    \vspace{-4mm}
    \caption{Compiler infrastructure:
    (a) Legacy gather code,
    (b) Performing tiling on the single loop,
    (c) Identifying and hoisting the indirect loads,
    and (d) Inserting \maa{} API calls.}
    \Description[\maa{} compiler infrastructure.]{\maa{} compiler infrastructure includes three passes to tile the loop, identify, hoist, and sink bulk data accesses before and after the loop, and convert bulk data accesses to \maa{} API calls.}
    \vspace{-3mm}
    \label{fig:compiler_trans_eg}
\end{figure}

\noindent\textbf{Compiler Transformation Overview.}
\maa{} compiler operates in three stages, as shown by Figure~\ref{fig:compiler_trans_eg}.
First, it converts the input C/C++ code into MLIR's target agnostic loop-level intermediate representation (\texttt{affine}, \texttt{scf}) using Polygeist~\cite{moses2021polygeist}, enabling loop-level transformations and analyses.
Next, it applies loop tiling to expose opportunities for bulk operations suitable for \maa{} offloading.

To identify indirect memory accesses, the compiler employs a Depth First Search (DFS) algorithm~\cite{swpref_1}.
It starts from loop induction variables and traverses the use-def chains to detect indirect access patterns within loops.
If indirect access is deemed legal for transformation, the compiler hoists indirect memory loads into \texttt{packed\_load} operations and sinks indirect store and RMW accesses into \texttt{packed\_store} and \texttt{packed\_RMW} operations from the innermost loop.
As illustrated in Figure \ref{fig:compiler_trans_eg} (b), the indirect access \texttt{A[B[i]]} is hoisted out of the loop into \texttt{packed\_load}, which is a structured operation that encodes the range of accesses and dependent memory instructions. \texttt{A[B[i]]} will be produced by a \texttt{packed\_load} using \texttt{enqueue} operation and consumed by \texttt{dequeue} operation in the original single \texttt{for} loop.
These packed operations serve as high-level abstractions representing bulk memory operations, equivalent to a target-agnostic interface for DAE~\cite{dae}.

During code generation, the compiler pattern-matches for packed operations and generates corresponding \maa{} APIs based on their encoded memory dependencies. Since \maa{} instructions execute asynchronously with the cores, the compiler inserts a \texttt{wait} API call when synchronization is required. This occurs before the core reads data from the scratchpad produced by load, ALU, or range fuser instructions, or when accessing global memory modified by \maa{}'s RMW or store operations.
Figure \ref{fig:compiler_trans_eg} (d) shows that \texttt{A[B[i]]} in \texttt{packed\_load} would be lowered to stream load and indirect load APIs mentioned in Section~\ref{subsec:programmint_API}.

\noindent\textbf{Legality.} 
Transforming and offloading indirect memory accesses to \maa{} requires that no core stores to the memory regions accessed by \maa{} within the loop body.
Additionally, \maa{} acceleration requires no data dependencies between different loop iterations.
We leverage MLIR's alias analysis to enforce these requirements.
For example, the \texttt{Gauss-Seidel} preconditioner used in multigrid solvers~\cite{AMG2023,hpcg} performs indirect loads from a data array while also storing to the array at different indices.
While this workload involves substantial indirect memory accesses, \maa{} cannot accelerate it because any aliasing between load and store indices could result in stale data if indirect loads are hoisted outside the loop.

\section{Evaluation Methodology}
\textbf{Benchmarks.}\maa{} is first evaluated under different cache conditions and DRAM access patterns using five microbenchmarks.
\footnotetext[1]{We follow prior work by evaluating bottom-up \texttt{BFS} and disabling buckets for \texttt{IS}. \maa{} can also accelerate data access in top-down \texttt{BFS} and bucket-based \texttt{IS} algorithms.}
Then, 12 workloads with indirect access patterns from five benchmark suites are evaluated for the main evaluation: (1) Conjugate Gradient (\texttt{CG}) of a $150K\times150K$ matrix and Integer Sort (\texttt{IS})\footnotemark[1] of $2^{25}$ keys from the NAS parallel benchmark suite~\cite{nas}, representative of the computational fluid dynamics applications.
(2) \texttt{GZ}, \texttt{GP}, \texttt{GZI}, and \texttt{GZI} workloads from the UME proxy~\cite{ume}, which perform a gradient computation over 2M zones and points of an unstructured mesh used in hydrodynamic simulations.
(3) Spatter benchmark~\cite{spatter} using an indirect access pattern collected using the methodology described in \cite{sheridan2024workflowsynthesis} from the \texttt{xRAGE} parallel multi-physics application~\cite{xrage}.
(4) Breadth-First Search (\texttt{BFS})\footnotemark[1], PageRank (\texttt{PR}), and Betweenness-Centrality (\texttt{BC}) graph algorithms from GAP benchmark suite~\cite{gap} using a uniform graph with $2^{20}$ to $2^{22}$ nodes and an average degree of 15.
(5) Two implementations of the Parallel Radix Join algorithms from the Hash-Join benchmark suite~\cite{hashjoin}, representative of the in-memory database workloads: histogram-based (\texttt{PRH})~\cite{histogram_prh} and bucket-chaining-based (\texttt{PRO})~\cite{bucketchaining_pro} using 2M tuples.

\textbf{Simulation Infrastructure.} \maa{} is integrated as a memory-mapped accelerator to the execution-based and event-driven Gem5 simulator~\cite{gem5} equipped with Ramulator2~\cite{ramulator2} as the backend memory simulator.
This setup accurately models both the core and memory access.
\maa{} is evaluated using a 4-core shared-memory system with configurations similar to the Intel Skylake architecture~\cite{skylake_hotchips,skylake_isscc} as shown by Table~\ref{tab:config}.
\textit{For a fair comparison with the baseline due to the scratchpad and area overhead of \maa{}, we increase the LLC size of the baseline by 2MB.}
A functional simulator for \maa{} APIs (see Section~\ref{subsec:programmint_API}) was developed to ensure the correctness of the implementations before simulation.
Correctness is re-verified with the Gem5 simulation.

\begin{table}[h]
  \footnotesize
  \centering
  \vspace{2mm}
  \caption{System Configuration}
  \vspace{-4mm}
  \setlength{\tabcolsep}{1pt} 
    \renewcommand{\arraystretch}{1.2} 
  \begin{tabular}{|c|c|}
    \hline 
    \textbf{Component} & \textbf{Configuration} \\
    \hline \hline
    \multirow{2}{*}{Core} & 4 8-wide cores, LQ: 72, SQ: 56, IQ: 50, ROB: 224, 180 Int Registers, \\
    & 168 Vector/Floating-Point Registers, 3.2GHz Frequency \\
    \hline
    L1I cache & 32KB, 8-way, 2 cycles latency, 16 MSHRs, Stride Prefetcher  \\
    \hline
    L1D cache & 32KB, 8-way, 4 cycles latency, 16 MSHRs, Stride Prefetcher  \\
    \hline
    L2 cache & 256KB, 4-way, 12 cycle latency, 32 MSHRs, Stride Prefetcher \\
    \hline
    \multirow{2}{*}{LLC} & \textbf{Baseline/DMP:} 10MB, 20-way,    \textbf{\maa{}}: 8MB, 16-way \\
    & 42 cycle latency, 256 MSHRs   \\
    \hline
    \multirow{3}{*}{Memory} & 2 Channels, DDR4 3200, 51.2GB/s Max BW, $t_{CK}$=625ps  \\
    & $t_{RP/RCD}$=12.5ns, $t_{CCDS/L}$=2.5/5.0ns, $t_{RTP}$=7.5ns, $t_{RAS}$=32.5ns \\
    & Request Buffer size: 32/channel, FR-FCFS scheduler \\
    \hline
    \hline
    \multirow{2}{*}{\maa{}} & 2MB Scratchpad w/ 4 ports: $32\times16K$ TILEs, 64$\times$8 Row Table slices \\
    & 32 Registers, 128 Request Table size, 16 ALU lanes, 256-entry TLB \\
    \hline
   \end{tabular}
  \label{tab:config}
\end{table}

\noindent\textbf{Indirect Prefetcher Modeling.}
We compare the performance and bandwidth utilization of \maa{} with the state-of-the-art indirect prefetcher, DMP~\cite{dmp}, using their public Gem5 artifact~\cite{dmp_artifact}.
We reproduced DMP's performance results using their best-performant workload, \texttt{IS}, and their \textit{single-core baseline} configuration with \textit{256KB last-level (L2) cache} size.
Then, we used the \textit{4-core} baseline configuration (Table~\ref{tab:config}) with \textit{10MB L3 cache} for comparison with \maa{}. Compared to the reported results~\cite{dmp}, we observed less performance improvement for DMP with larger caches and more cores in our baseline configuration.

\noindent\textbf{Power and Area Modeling.}
\maa{} was implemented in RTL and  all components were synthesized with Synopsys Design Compiler using 28nm TSMC library.
BCAM area and power for the Row Table is evaluated in 28nm FDSOI technology~\cite{bcam}.
The baseline Skylake core area is evaluated using die shots~\cite{die_shot}.
To compare \maa{} area with the baseline core, all area and power numbers are scaled to 14nm using equations of~\cite{techscale}.
\begin{figure*}[thb]
    \centering
    \includegraphics[width=0.99\textwidth]{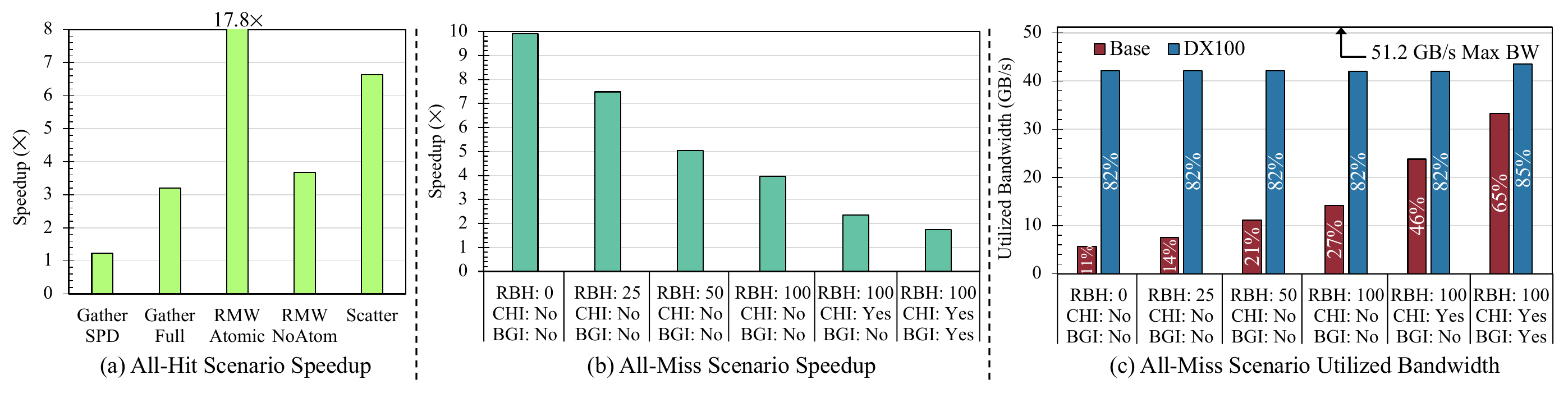}
    \vspace{-2mm}
    \caption{(a) \maa{} speedup for different access types under 100\% L1 cache hit rate, (b) \maa{} speedup and (c) utilized bandwidth for Gather-Full kernel w.r.t. row buffer hit rate (RBH), channel interleaving (CHI), and bank-group interleaving (BGI).}
    \Description[Performance improvement in All-Hit and All-Miss microbenchmarks.]{(a) In the All-Hit case, all accesses hit in the L1 cache; \maa{} improves significantly the performance for read-modify-write microbenchmark as it eliminates fine-grain atomic operations. (b) and (c) In the All-Miss case, all accesses go to DRAM; \maa{} improves gather performance by up to 9.9$\times$, depending on baseline index characteristics like row-buffer hit rate, channel, and bank-group interleaving.}
    \label{fig:microbenchmark_results}
    \vspace{-3mm}
\end{figure*}

\section{Results}

\subsection{Microbenchmarks}~\label{subsec:results_microbenchmark}
To evaluate \maa{} benefits under certain index distributions, we implement five microbenchmarks with Gather, Scatter, and RMW access patterns under two scenarios.


 \noindent\textbf{Scenario 1: All-Hits.} 
In this scenario, we warm up all caches and use a streaming index distribution (\texttt{B[i] = i}), enabling cores to fetch all data directly from the L1 cache, thus there are no benefits from memory access reordering for this scenario.
This setup highlights the ability to accelerate address calculation operations and reduce the dynamic instruction count. 
Figure~\ref{fig:microbenchmark_results} (a) illustrates the performance gains of \maa{} over the baseline.

In \textit{Gather-SPD}, we only offload the gather operation (\texttt{p\_A[i] = A[B[i]]}) to \maa{}.
The packed array (\texttt{p\_A[i]}) is loaded from the scratchpad (SPD) to the cores.
While \maa{} reduces the dynamic instruction by 2.9$\times$, loading the packed array from the higher latency SPD limits the overall performance speedup to 1.2$\times$. The SPD latency impacts performance minimally when we move to realistic benchmarks with higher cache misses and memory accesses. 

In the \textit{Gather-Full} experiment, the entire kernel (\texttt{C[i]=A[B[i]]}) is offloaded to \maa{} by using a streaming store operation on \texttt{C[i]}, achieving a 3.2$\times$ performance gain by reducing CPU core's dynamic instructions from 870K in the baseline to just 273 with \maa{}.

For RMW operations, the baseline implementations rely on atomic accesses to ensure correctness, which is ${\approx}4.8\times$ slower than non-atomic operations because of the memory fences~\cite{free_atomics} and cacheline or bus locking mechanisms~\cite{intel_software_manual}.
As a shared accelerator and the sole writer to indirect memory regions (Section~\ref{subsec:compiler}), \maa{} eliminates fine-grain atomic operations.
To assess \maa{} benefits for RMW (\texttt{A[B[i]] += C[i]}), we compare it to two baseline implementations: one using atomic operations (\textit{RMW-Atomic}) and another ignoring correctness (\textit{RMW-NoAtom}).
\maa{} outperforms these baselines by 17.8$\times$ and 3.7$\times$, respectively, demonstrating high efficiency for common RMW operations (Table~\ref{tab:pattern}).

\textit{Scatter} operations (\texttt{A[B[i]] = C[i]}) cannot be parallelized in a shared memory system even with atomic operations as reordering the indirect stores results in WAW data hazards and compromises correctness.
Therefore, we use a single-core configuration with a 4MB LLC for the baseline and a 2MB LLC for \maa{}.
In this configuration, \maa{} improves the scatter performance by 6.6$\times$. This speedup is 2$\times$ higher than Gather-Full primarily because the baseline is restricted to one core due to data hazards.

\noindent\textbf{Scenario 2: All-Misses.}
To evaluate the memory bandwidth improvements of \maa{}, we analyze an All-Miss scenario using the \textit{Gather-Full} microbenchmark.
In this setup, all indirect accesses miss the cache, requiring data to be fetched directly from DRAM.
We generate a constant set of 64K unique \texttt{B[i]} indices to evenly distribute the indirect \texttt{A[B[i]]} words across 16 rows in all banks, bank groups, and channels.
These unique indices are then reordered to artificially create various \textbf{row-buffer hit rates (RBH)} (0\%-100\%), \textbf{channel interleaving (CHI)}, and \textbf{bank-group interleaving (BGI)} patterns between consecutive indices.

Figures~\ref{fig:microbenchmark_results} (b) and (c) present the performance and bandwidth utilization of \maa{} compared to the baseline. The row-buffer hit rates (RBH) and memory-level parallelism (CHI, BGI) increase in the benchmark access patterns from left (worst) to right (best); \maa{} performance improvements decrease from left to right, with a maximum improvement of 9.9$\times$ with the worst-case index pattern.

Several key observations are evident.
\textit{First}, \maa{} consistently achieves high bandwidth utilization, ranging from 82\% to 85\%, regardless of the order of the indices in the input dataset.
This is due to \maa{}'s ability to reorder memory addresses.
In contrast, the baseline relies on DRAM memory controllers, which have limited reordering capabilities due to limited visibility into future memory requests.
\textit{Second}, even with the best-case index ordering for the baseline with 100\% RBH, CHI, and BGI (right-most bar), it achieves only 65\% of the maximum 51.2GB/s bandwidth because of its limited memory-level parallelism.
In this configuration, \maa{} achieves a $1.7\times$ speedup, primarily because of its higher memory access rate.
\textit{Third}, with no BGI (second right-most bar), the theoretical bandwidth utilization drops to half as the column-to-column read is constrained by $t_{CCDL}$.
Here, the baseline utilizes 46\% of DRAM bandwidth.
When CHI is further eliminated, the baseline channel utilization drops from 46\% to 27\%, effectively leaving one channel idle.
Finally, as the row-buffer hit rate decreases from 100\% to 0\% (left-most bars), the baseline's bandwidth utilization drops by 2.5$\times$.

\raggedbottom

\subsection{Performance Analysis}~\label{subsec:performance_analysis}
\begin{figure}[h]
    \centering
    \includegraphics[width=0.85\linewidth]{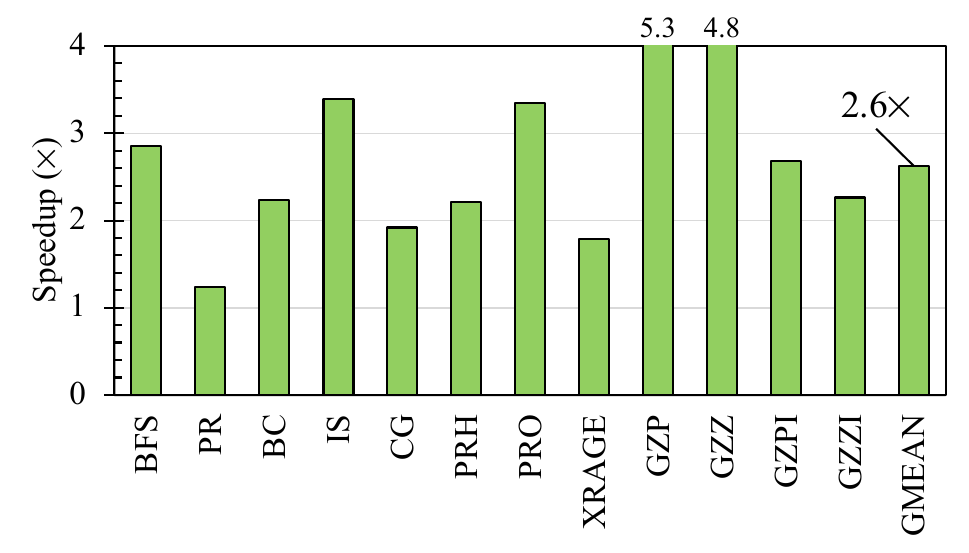}
    \vspace{-4mm}
    \caption{\maa{} speedup for different workloads.} 
    \Description[Performance speed-up across 12 benchmarks.]{\maa{} achieves a geometric mean speed-up of 2.6$\times$ across 12 evaluated benchmarks.}
    \label{fig:speedup_results}
\end{figure}

Figure~\ref{fig:speedup_results} shows that \maa{} achieves a geometric mean speedup of 2.6$\times$ compared to the baseline across 12 irregular workloads.
This speedup is due to significantly higher bandwidth utilization, instruction reduction, and cache miss reductions.
In the following paragraphs, we discuss each metric in detail.

\begin{figure*}[thb]
    \centering
    \includegraphics[width=0.95\textwidth]{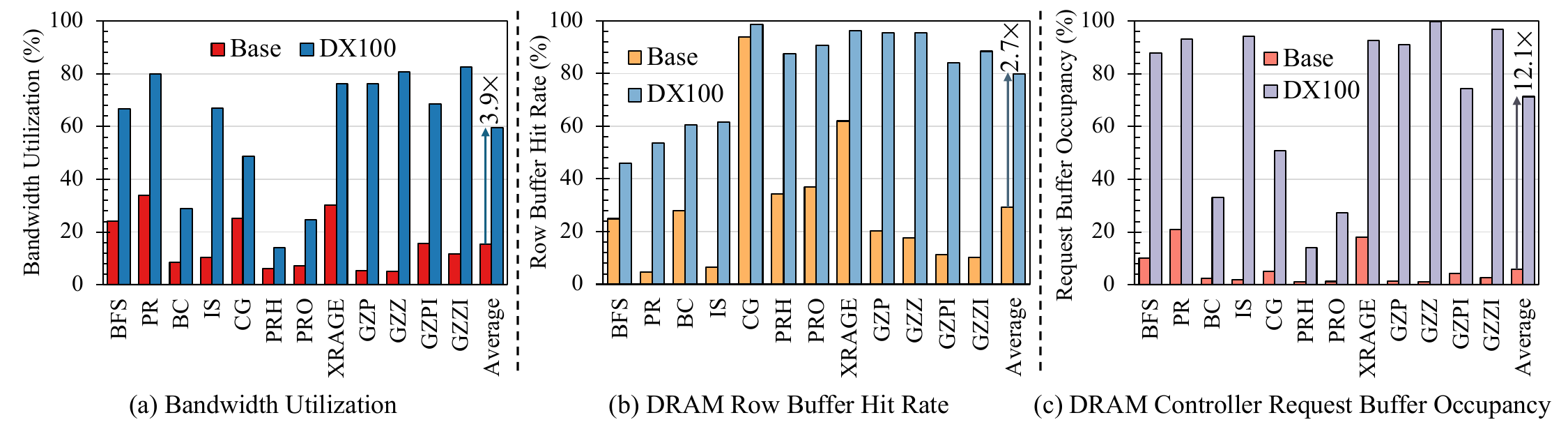}
    \vspace{-5mm}
    \caption{(a) Bandwidth utilization, (b) row buffer hit rate, and (c) request buffer occupancy of baseline vs. \maa{}.}
    \Description[Memory bandwidth improvement across 12 benchmarks.]{\maa{} improves average memory bandwidth utilization by 3.9$\times$ due to a 2.7$\times$ increase in row-buffer hit rate because of the reordering and interleaving techniques and a 12.1$\times$ increase in DRAM access rate via direct memory controller access.}
    \label{fig:bandwidth_results}
    \vspace{-4mm}
\end{figure*}

\noindent\textbf{Higher Bandwidth Utilization.}  
Figure~\ref{fig:bandwidth_results} (a) compares the bandwidth utilization of \maa{} with the baseline, showing an average improvement of 3.9$\times$.
The bandwidth improvement is particularly higher in workloads with significant indirect memory accesses, as \maa{} enhances the efficiency of these accesses through reordering, coalescing, and interleaving techniques. 
For instance, the \textit{Integer Sort} (\texttt{IS}) kernel from the \texttt{NAS} benchmark suite involves indirect indexing to a large key array containing $2^{25}$ elements.
Here, indirect accesses consume a significant portion of the memory bandwidth, and \maa{} improves the bandwidth utilization by 6.5$\times$.
Conversely, the Conjugate Gradient (\texttt{CG}) kernel from the same benchmark suite operates on a sparse matrix format, where most memory accesses involve streaming to the matrix, with relatively fewer indirect accesses targeting a vector.
Consequently, \maa{} achieves a lower improvement of 1.9$\times$ in bandwidth utilization for this kernel.

\maa{} improvement of memory bandwidth utilization for indirect accesses is primarily attributed to two key factors:

First, \textit{\maa{} significantly improves the DRAM row-buffer hit rate by reordering memory accesses}, achieving an average increase of 2.7$\times$, as shown in Figure~\ref{fig:bandwidth_results} (b).
The row-buffer hit rate is defined as the proportion of memory accesses that find their target column in the row-buffer.
\maa{} substantially enhances the row-buffer hit rate for UME kernels, \textit{i.e.,} \texttt{GZP}, \texttt{GZZ}, \texttt{GZPI}, and \texttt{GZZI}.
Analyzing the UME dataset, consisting of 2M data points, reveals an average index distance ($abs(i - B[i])$) of 85K elements, indicating limited spatial locality in the data.
Despite this, \maa{}'s reordering technique within a tile of 16K elements successfully grouped 7.6 column accesses per DRAM row, allowing the indirect access unit to issue these accesses consecutively.
This optimization improves the row-buffer hit rate of UME kernels by 6.1$\times$, from 15\% to 91\%.

Second, \textit{\maa{} increases the memory-level parallelism and access rate}.
To evaluate this claim, we measure the request buffer occupancy, defined as the ratio of the average number of memory accesses in the DRAM memory controller's request buffer to the maximum buffer size (32).
Occupancy is a critical factor for DRAM controllers, enabling the interleaving of commands across different banks and hiding the latency of \texttt{PRE} and \texttt{ACT} commands for one bank behind the latency of reads and writes to another bank.
\maa{} improves the request buffer occupancy of the baseline by 12.1$\times$ because it breaks the dependency between index loads and indirect accesses, enabling bulk memory access.
Furthermore, \maa{} is placed near the memory controllers, bypassing the structural limitations in the core and memory system.
On the other hand, the baseline suffers from limited outstanding memory accesses because only 54 instructions in a 224-entry ROB are loads and stores across all cycles of all benchmarks.
These load and store instructions generate a chain of dependencies with prior memory accesses and address calculation instructions~\cite{droplet}.
On average, 19 outstanding loads in the LQ are issued to the memory system at any given cycle.
From these, only 10 accesses target data arrays, and the rest are stack accesses or register spills and fills, which often hit the L1 cache.
Finally, many word accesses are coalesced in the MSHRs and hit in the cache.
Together, these factors reduce the request buffer occupancy of the baseline to an average of only 2 memory accesses, highlighting the substantial improvements enabled by \maa{}.

\begin{figure}[h]
    \centering
    \includegraphics[width=0.98\linewidth]{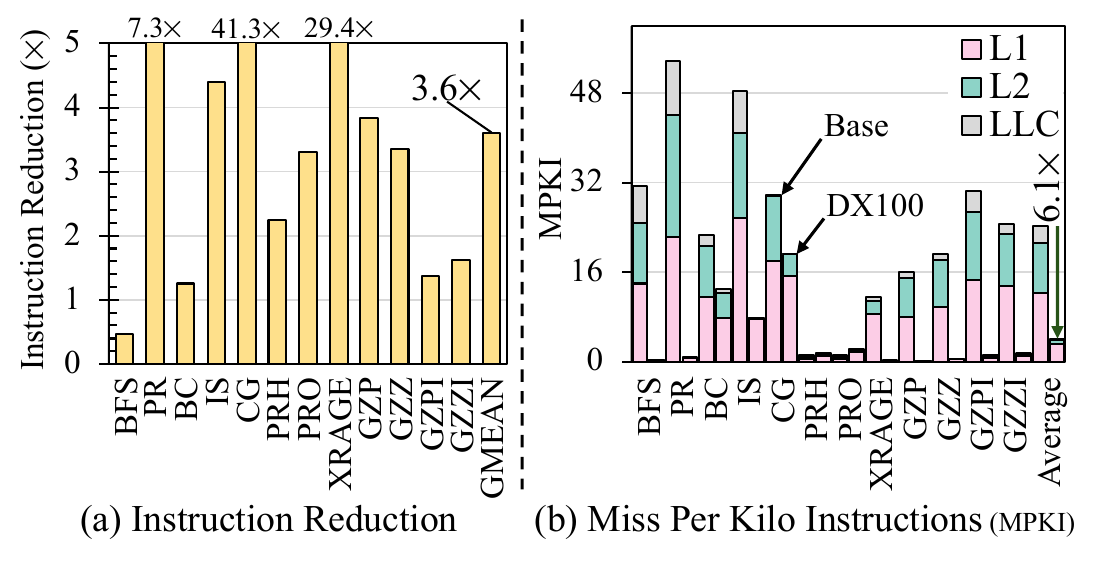}
    \vspace{-4mm}
    \caption{\maa{} (a) instruction and (b) MPKI reduction.}
    \Description[Instruction count reduction and cache performance improvement.]{(a) \maa{} reduces dynamic instruction count by a geometric mean of 3.6$\times$ through address computation acceleration. (b) Cache MPKI is reduced by an average of 6.1$\times$ by avoiding cache pollution from index and indirect data accesses.}
    \vspace{-4mm}
    \label{fig:instr_mpki_results}
\end{figure}

\noindent\textbf{Instruction and Cache Miss Reduction.}
Figure~\ref{fig:instr_mpki_results} (a) shows that \maa{} reduces the number of core instructions by a geometric mean of 3.6$\times$, which can significantly improve CPU core energy consumption. The reduction in dynamic instruction count is achieved primarily by accelerating indirect address calculation operations.
However, \texttt{BFS} instruction count slightly increases in \maa{} implementation.
This is due to the use of OpenMP \texttt{critical} primitives for synchronization between cores accessing \maa{}, which rely on spinning locks.
These locks increase the instruction count while waiting for another core to complete its \maa{} instructions.

Despite fewer instructions, Figure~\ref{fig:instr_mpki_results} (b) shows that \maa{} significantly reduces the cache MPKI, measured across the whole workload execution.
This improvement is because indirect accesses in the baseline often poorly utilize cache lines, leading to cache pollution and wasted capacity.
In contrast, \maa{} directly injects these accesses into memory, packs them, and writes the packed data to the Scratchpad.
Cores access this packed data in a streaming manner from the Scratchpad, preserving cache space.
Additionally, intermediate index arrays are accessed exclusively by \maa{}, freeing up private cache space for cores.
As a result, \maa{} effectively increases the usable cache capacity for critical core accesses.

\subsection{Comparison with Indirect Prefetcher}~\label{subsec:prefetcher}
\begin{figure}[b]
    \centering
    \includegraphics[width=0.98\linewidth]{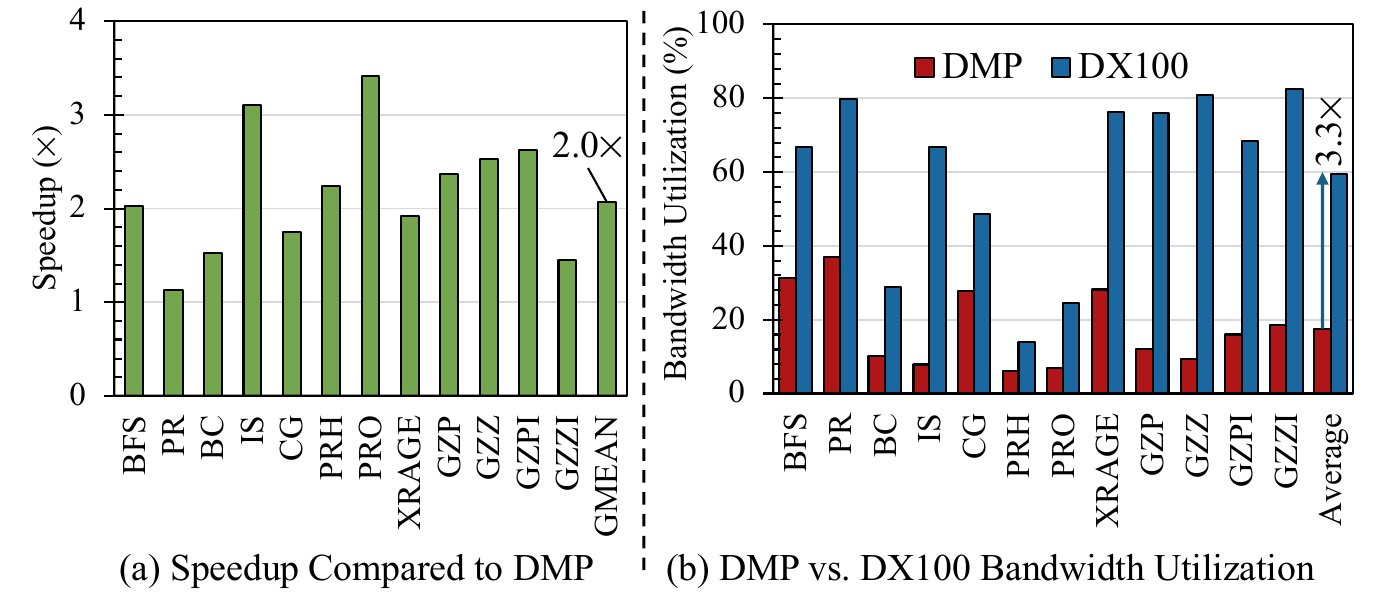}
    \vspace{-2mm}
    \caption{(a) Speedup and (b) Bandwidth Utilization of \maa{} compared to DMP~\cite{dmp}.}
    \Description[Comparison to state-of-the-art indirect prefetcher.]{(a) \maa{} outperforms DMP with a 2.0$\times$ geometric mean speed-up. (b) \maa{} achieves 3.3$\times$ higher bandwidth utilization on average.}
    \label{fig:dmp_results}
\end{figure}

Figure~\ref{fig:dmp_results} (a) shows that \maa{} outperforms the state-of-the-art indirect prefetcher, DMP~\cite{dmp}, achieving a 2.0$\times$ geometric mean speedup. This is due to \maa{}'s 3.3$\times$ higher bandwidth utilization, as depicted in Figure~\ref{fig:dmp_results} (b). While DMP improves baseline bandwidth by increasing memory access rates, it \textit{does not reorder} memory accesses and primarily relies on the memory controllers for DRAM command optimization. In contrast, \maa{}'s bulk accesses surpass the limited future visibility of memory controllers, enabling better bandwidth utilization with effective reordering.

\maa{} reduces dynamic instructions and improves energy consumption of CPU cores, unlike indirect prefetchers that leave the instruction footprint unchanged. Indirect prefetchers primarily lower memory access latency by improving cache hit rates; for instance, DMP reduces the average memory access latency by an average of 1.4$\times$ over the baseline. However, accurate prefetching is challenging for irregular workloads with conditional accesses. Many of the evaluated workloads involve conditional accesses (Table~\ref{tab:pattern}) that are difficult for prefetchers to handle effectively. Prefetching untaken loop iterations degrades performance by polluting the cache, especially with low cache line utilization of indirect accesses. In contrast, \maa{} ensures 100\% accurate fetching through a programmable interface supporting ALU operations and conditional accesses.
\vspace{-2mm}
\subsection{Sensitivity Analysis}~\label{subsec:tile_size_results}
\begin{figure}[h]
    \centering
    \includegraphics[width=0.98\linewidth]{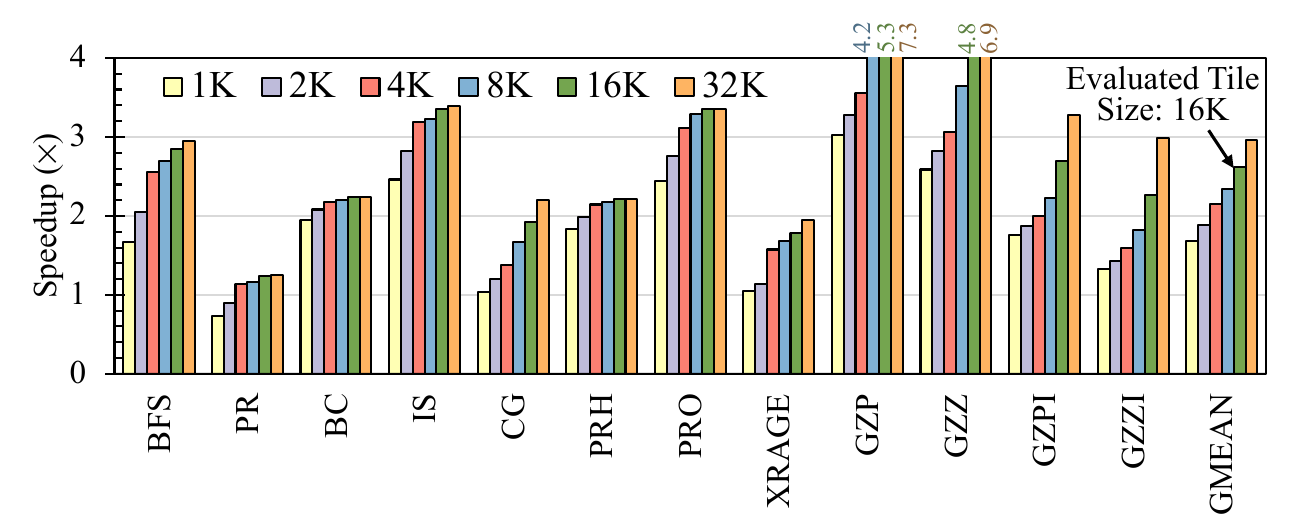}
    \vspace{-2mm}
    \caption{Performance sensitivity to the tile size.}
    \Description[\maa{} speed-up versus tile size.]{As the tile size increases from 1K to 32K, \maa{}'s geometric mean speed-up improves from 1.7$\times$ to 2.9$\times$ over the baseline.}
    \label{fig:tile_size_results}
\end{figure}

Figure~\ref{fig:tile_size_results} shows that increasing tile size from 1K to 32K elements improves the performance from 1.7$\times$ to 2.9$\times$ over the baseline. This speedup arises from two primary factors.
\textit{First,} \maa{} reduces memory accesses by coalescing redundant addresses within larger tiles; 32K tile size reduces memory accesses by a geometric mean of $1.4\times$ compared to a 1K tile size.
\textit{Second,} \maa{} increases memory bandwidth utilization by 25\% when enlarging the tile size from 1K to 32K, primarily due to a 27\% higher row-buffer hit rate with a 32K tile size.
Notably, increasing the tile size has minimal impact on the memory access rate (DRAM controller occupancy), as this is primarily determined by the number of outstanding memory accesses, which remains unchanged by the tile size.

\subsection{Area and Power Analysis}

Table~\ref{tab:area} provides a detailed area and power breakdown for all \maa{} components, synthesized using a $28~nm$ library.
The area and power are dominated by the Scratchpad, which contains 32 tiles of 16K elements, facilitating the communication between \maa{} components and cores.
Using die shots of a 4-core Skylake processor~\cite{die_shot} fabricated in a 14nm technology node, we estimate the area of a core to be approximately $10.1mm^2$.
Of this, around $2.3mm^2$ is occupied by a 2MB cache slice, which includes data arrays, tag arrays, and coherency directory.
Applying the scaling factors from~\cite{techscale}, we estimate \maa{} area in 14nm to be approximately $1.5mm^2$ which results in a 3.7\% overhead for the processor as \maa{} is shared between four cores. Since \maa{} area is comparable to the area of a single cache slice in the baseline architecture, \textit{all evaluations in this paper use a baseline with a 2MB larger LLC} (see Table~\ref{tab:config}).
Our evaluation shows around 777$mW$ power consumption for \maa{}.

\begin{table}[h]
  \centering
  \footnotesize
  \vspace{1mm}
  \caption{\maa{} Area and Power Analysis in $28~nm$}
  \vspace{-3mm}
  \begin{tabular}{|c|c|c|}
    \hline
    \textbf{Module} & \textbf{Area ($mm^2$)} & \textbf{Power ($mW$)} \\
    \hline
    \hline
    Range Fuser & 0.001 & 0.26 \\
    \hline
    ALU & 0.095 & 74.83 \\
    \hline
    Stream Access & 0.012 & 6.03 \\
    \hline
    Indirect Access & 0.323 & 83.70 \\
    \hline
    Controller & 0.002 & 0.43 \\
    \hline
    Interface & 0.045 & 30.0 \\
    \hline
    Coherency Agent & 0.010 & 3.12 \\
    \hline
    Register File & 0.005 & 1.56 \\
    \hline
    Scratchpad & 3.566 & 577.03 \\
    \hline
    \hline
    Total & 4.061 & 777.17 \\
    \hline
   \end{tabular}
  \label{tab:area}
\end{table}

\subsection{Scalability Discussion}~\label{subsec:scalability}
Our evaluation in prior sections focused on a system with four cores and two DDR4-3200 memory channels. When scaling the System-on-Chip (SoC) with more cores and memory channels, multiple instances of \maa{} can be integrated to utilize the increased memory bandwidth.
We explored two approaches for integrating multiple \maa{} instances as discussed next.

\noindent\emph{(1) Address range partitioning:} Assigns a \maa{} instance to a specific memory address range or region. The programmer or compiler ensures that an entire data structure or array is allocated within a single region, guaranteeing exclusive access by a single \maa{} instance. This exclusivity ensures the correctness of RMW operations without additional design complexity. However, if interacting arrays are assigned to different regions, scratchpad tile transfers between \maa{} instances become necessary. While allocating all interacting data structures to the same region can eliminate these transfers, it may also lead to load imbalance among \maa{} instances limiting scalability.

\noindent\emph{(2) Core multiplexing:} Assigns multiple cores to a \maa{} instance for acceleration, ensuring that all \maa{} instructions from a given core are directed to a specific instance. This design places \maa{} instances closer to their associated cores within the SoC, reducing latency for core-scratchpad accesses. To maintain correctness for RMW operations, only one instance must have exclusive write access to an indirect array. This is enforced through a \textit{coarse-grained region-based coherence protocol}~\cite{region_scout}, which upholds the Single-Writer-Multiple-Reader (SWMR) invariant~\cite{coherency_premier} among instances. \maa{} receives address ranges from page table entry transfers and maintains them in the TLB (Section~\ref{subsec:interface_CA}).
When \maa{} dispatches memory access instructions, the entire address range covering the array (\texttt{\&A[0]} to \texttt{\&A[SIZE]}) is treated as a coherence region. The Coherency Agent verifies the coherence state and issues a coherence request if the region is not in the required state. Memory access instructions are issued once the appropriate state is acquired. During execution, the coherence region is locked to prevent other instances from obtaining write permission. This protocol operates independently of the primary multicore coherence mechanism, as \maa{} assumes that cores do not simultaneously modify arrays (Section~\ref{subsec:compiler} -- Legality).

Due to the overhead of inter \maa{} tile transfers and load imbalance limitations of the address range partitioning, we implement the second approach, core multiplexing, in Gem5.
We study the performance scalability by doubling the cores, LLC size, memory channels, and the input dataset size of the benchmarks.
We evaluate two \maa{} configurations: one with a single instance but a 4MB scratchpad shared by eight cores, and another with two instances each with a 2MB scratchpad and shared by four cores.

\begin{figure}[h]
    \centering
    \includegraphics[width=0.98\linewidth]{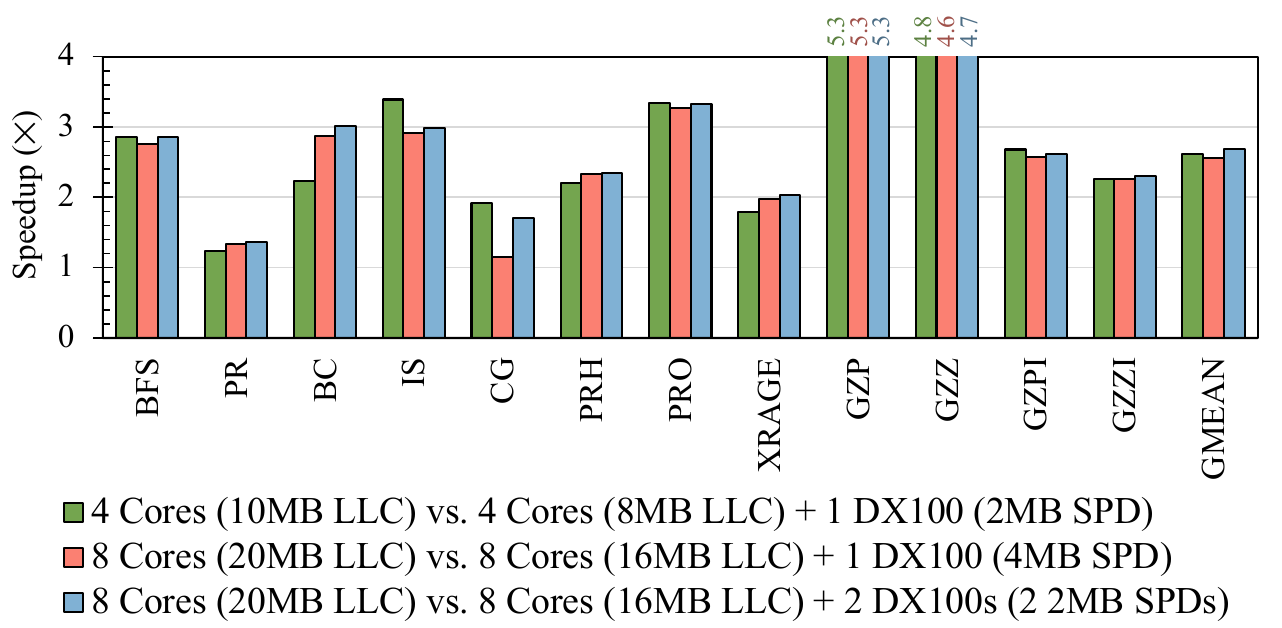}
    \vspace{-2mm}
    \caption{Performance improvement scalability using different number of cores and \maa{} instances.}
    \Description[Scalability with different number of cores, memory channels, and \maa{} instances.]{When increasing the number of cores from 4 to 8 and number of memory channels from 2 to 4, performance slightly drops from 2.6$\times$ to 2.5$\times$ using one \maa{} instance. Adding a second instance increases speed-up to 2.7$\times$.}
    \label{fig:scalability_results}
    \vspace{-2mm}
\end{figure}

Figure~\ref{fig:scalability_results} presents the performance scalability results normalized to baselines with the same core count, highlighting two key takeaways. First, \maa{} sustains its performance advantage over the baseline system as the core count scales to eight, achieving a geometric mean speedup of 2.6$\times$ with four cores (green bars) and 2.5$\times$ with eight cores (orange bars). A single \maa{} instance utilizes 60\% of the available bandwidth across two memory channels (51.2 GB/s peak) and 54\% across four memory channels (102.4 GB/s peak).
Second, the core multiplexing approach does not introduce significant bottlenecks from additional coarse-grained region-based coherence messages and checks for multiple instances of \maa{}. Despite this overhead, integrating two \maa{} instances even increases speedup to 2.7$\times$ (blue bars) due to the use of additional functional units.

\section{Related Work}

Efficient data access in irregular applications is a longstanding challenge due to the prevalence of indirect memory accesses.
Here, we discuss many approaches to mitigate the memory access bottleneck.

\textbf{Prefetcher.} 
Software prefetching~\cite{swpref_1,APT_GET,resource_sw_pref,spaid_sw_pref, cmpPrefetch1991,shared_sw_pref} aims to hide memory latency by inserting prefetch instructions for data likely to be accessed soon. 
These techniques leverage static analysis, runtime profiling, or programmer annotations to detect indirect loads. 
Hardware prefetchers~\cite{dmp,imp,hw_prefetch_1,hw_prefetch_bingo,hw_prefetch_complex, temp_prefetcher, correlated_prefetcher, epoch_prefetcher, tag_prefetcher, linear_prefetcher, markov_prefetcher, best_offset_prefetcher, meta_prefetcher, without_meta_prefetcher,gpu_prefetch,ssd_prefetch, dual_spatial} predict future accesses at runtime by exploiting spatial and temporal locality. 
Hybrid approaches~\cite{prodigy,Learning_pref,event_pref, graph_pref} combine compiler or programmer annotations with hardware support to improve prefetching accuracy.
Runahead execution~\cite{runahead_efficient,runahead_ooo,runahead_org,vector_runahead,runahead_percise, runahead_scalar,decouple_runahead1, decouple_runahead2} allows the processor to speculatively execute instructions when it is stalled due to long-latency memory accesses.
Prefetching effectively reduces average memory access latency but provides limited improvements to memory bandwidth utilization.
In contrast, \maa{} enhances the memory bandwidth through reordering, coalescing, and interleaving.
Prefetchers do not impact the instruction footprint, while \maa{} accelerates address calculation operations.
Moreover, prefetchers suffer from low accuracy with conditional memory accesses, leading to cache pollution and wasting the cache capacity.
\maa{} supports conditional operations and injects the indirect accesses to the DRAM, improving cache line utilization and reducing the cache misses for other critical data accesses of the core.

\textbf{Fetcher Units}, inspired by Decoupled Access-Execute paradigm~\cite{dae}, aim to improve performance by separating and offloading programs into independent access and execute cores. This decoupling allows memory accesses and computations to be processed concurrently on dedicated processors, hiding memory latency. Fetcher units serve as the access core to perform specific memory accesses and feed the data to the processor to avoid recomputation.
Specialized fetchers have been proposed for various domains. HATS~\cite{HATS} performs locality-aware graph traversals to reduce data movement. SPU~\cite{spu}, TMU~\cite{TMU} and MAPLE~\cite{MAPLE} can support traversal of sparse tensors and processing of sparse linear algebra. Widx~\cite{widx} accelerates hash indexing operations. SQRL~\cite{sqrl} handles traversals of hash tables and tree data structures. General-purpose fetcher units such as SpZip~\cite{spzip} and Terminus~\cite{lee2024terminus} support a wider range of applications and exploit fine-grained pipelining to hide the latency of indirect memory access chains. SpZip incorporates data compression to reduce memory traffic, while Terminus supports fine-grained updates for recursive data structures like trees and linked lists.
However, fetchers provide insufficient visibility into future memory accesses, which hinders effective memory access reordering and coalescing.
\maa{} differs by providing a programmable accelerator capable of improving memory bandwidth utilization for bulk memory accesses by reordering indices. 

\textbf{Specialized Accelerators} have been proposed to offload both data access and computation for irregular workloads. Near-memory processing architectures~\cite{RECNMP,HB-PNM,linkedlist_NMP,Tesseract,spacea,emc} and Near-cache processing architectures~\cite{tako,HAT,phi,cobra,sortcache,ASA} place computation close to memory or cache to reduce the memory access latency that is imposed by data movement.
While \maa{} also reduces the data transfer overhead of indirection, its primary benefit is from enhancing the memory bandwidth utilization.
Additionally, some accelerators~\cite{graphicionado,extensor,outerspace,Capstan,aurochs,innersp,flexagon,wang2024data} employ customized architectures and reconfigurable data-flow designs to handle irregular data access patterns and exploit parallelism effectively.
These specialized accelerators typically support a narrow set of applications and require substantially redesigning existing algorithms or data structures to fit their specific hardware models. In contrast, \maa{} offers a general-purpose solution that supports multiple degrees of indirection, diverse loop patterns, and various access types without significantly changing existing algorithms.

\textbf{Memory access reordering.}
DRAM memory controllers~\cite{virtualchannel,frfcfs,mutlu2008parallelism,mcrl,streammc,historymc,brustmc} have been designed to reorder requests to minimize page misses, bank conflicts, and improve row-buffer hit rates. However, they can schedule requests only within a limited window of visibility—typically around 48 cache lines per memory channel~\cite{ivyServer}—which restricts reordering opportunities. 
To exploit reordering beyond this window, software solutions like Milk~\cite{milk} and Propagation Blocking~\cite{propblocking} first collect random indirect memory accesses into cache-fitting batches and then reorder them into efficient sequential DRAM accesses. Propagation Blocking specifically accelerates PageRank, while Milk is a compiler that supports a broad set of applications with programmer annotations. However, these software solutions introduce timing and space overhead due to the need to introduce intermediate data structures, which can pollute the cache and sacrifice temporal locality. 
Hardware approaches like PHI~\cite{phi} and COBRA~\cite{cobra} modify the cache hierarchy to defer writes or read-modify-write operations with poor spatial locality, batching them to achieve sequential main memory accesses. Crescent~\cite{crescent} co-designs an approximate neighbor search algorithm and a hardware accelerator to convert irregular DRAM accesses to sequential DRAM accesses for deep point cloud analytics workloads. GSDRAM~\cite{seshadri2015gather} supports access with non-unit strides on commodity DRAM to improve bandwidth utilization but does not optimize indirect accesses.
In contrast, \maa{} offers several advantages over these approaches. Compared to memory controllers, \maa{} provides a much larger window for reordering memory requests, enabling more effective memory bandwidth utilization. Unlike software solutions, \maa{} operates at the hardware level without introducing space overhead or causing cache pollution from intermediate data structures. Moreover, compared to existing hardware solutions, \maa{} can support complex and general-purpose indirect memory access patterns.

\section{Conclusion}

\maa{} is a programmable data access accelerator designed to optimize memory bandwidth utilization for irregular applications using two ideas.
First, \maa{} is placed near the memory controllers, allowing it to inject memory accesses at a higher rate and bypass the core and memory system limitations.
Second, \maa{} employs three key techniques over bulk indices—\textit{reordering}, \textit{coalescing}, and \textit{interleaving}—to improve DRAM row-buffer hit rate, reduce memory accesses, and optimize the bandwidth of DRAM channels and bank-groups.
Using \maa{}'s general-purpose ISA with 8 instructions, we accelerate 12 benchmarks accross five benchmark suites, and observe 2.6$\times$ performance improvement, 3.9$\times$ higher bandwidth utilization, and 3.6$\times$ instruction reduction.

\bibliographystyle{ACM-Reference-Format}
\bibliography{main}

\clearpage

\appendix
\section{Artifact Appendix}

\subsection{Abstract}

This appendix provides a step-by-step guide to reproduce the main results shown in Figures~\ref{fig:speedup_results}, \ref{fig:bandwidth_results}, and \ref{fig:instr_mpki_results}. 
It includes instructions for cloning the GitHub repository, building the simulator and benchmarks, running simulations, processing the results, and plotting the charts.
The simulation environment uses the event-driven gem5~\cite{gem5} simulator for modeling the CPU core and \maa{}, and Ramulator2~\cite{ramulator2} for accurately simulating DRAM access bandwidth and latency.
Simulation infrastructure, benchmarks, automation scripts, expected results, and detailed instructions are available in our \href{https://github.com/arkhadem/DX100}{\blue{GitHub repository}}.

\subsection{Artifact check-list (meta-information)}

{\small
\begin{itemize}
  \item {\bf Program:} NAS~\cite{nas}, GAP~\cite{gap}, Hash-Join~\cite{hashjoin}, UME~\cite{ume}, and Spatter~\cite{spatter} benchmarks are included in the \href{https://github.com/arkhadem/DX100/tree/main/benchmarks}{\blue{\texttt{benchmarks} directory of the GitHub repository}}.
  \item {\bf Compilation:} \texttt{g++-12}, \texttt{clang++-15}, \texttt{SCons 3.0+}, \texttt{Python 3.6+}.
  \item {\bf Data set:} Benchmarks use synthetic datasets, except for the \texttt{XRAGE} dataset used by the Spatter benchmark, that is \href{https://web.eecs.umich.edu/~arkhadem/projects/xrage.tar.gz}{\blue{available online}}.
  \item {\bf Metrics:} Execution time, DRAM bandwidth, row buffer hit rate, request buffer occupancy, instruction count, and cache Misses Per Kilo Instructions (MPKI).
  \item {\bf Output:} Charts plotted using automation scripts available in the \href{https://github.com/arkhadem/DX100/tree/main/results}{\blue{\texttt{results} directory of the GitHub repository}}.
  \item {\bf Experiments:} All necessary steps are in the \href{https://github.com/arkhadem/DX100/blob/main/README.md}{\blue{\texttt{README} file}}. A Docker image is provided on \href{https://hub.docker.com/repository/docker/arkhadem95/dx100/general}{\blue{Docker Hub}} to simplify setup and evaluation.
  \item {\bf How much disk space required?:} 6GB for simulation infrastructure and benchmarks, plus 20GB for results.
  \item {\bf How much memory required?:} 35GB for each simulation.
  \item {\bf How much time is needed to prepare workflow?:} 35 minutes.
  \item {\bf How much time is needed to complete experiments:} Around 84 hours for serial execution (one simulation at a time using 35GB memory), or 24 hours when running four simulations in parallel (140GB memory).
  \item {\bf Publicly available?:} Both \href{https://github.com/arkhadem/DX100}{\blue{GitHub}} and \href{https://doi.org/10.5281/zenodo.15103396}{\blue{Zenodo}}.
  \item {\bf Code licenses?:} \href{https://github.com/arkhadem/DX100/blob/main/LICENSE}{\blue{MIT License}}.
  \item {\bf Workflow automation framework used?:} Automation scripts are available on the \href{https://github.com/arkhadem/DX100/tree/main/scripts}{\blue{\texttt{scripts} directory}}. 
\end{itemize}
}

\subsection{Description}

This artifact has several dependencies that are mentioned on the \href{https://github.com/arkhadem/DX100/blob/main/README.md}{\blue{\texttt{README} file}}. To simplify the evaluation process, we provide a Docker image with all required dependencies pre-installed. We recommend using the Docker container to minimize setup time and ensure a consistent environment. Use the following script to pull the image and create a container:

\begin{lstlisting}
docker pull arkhadem95/dx100:latest
docker run -it --name dx100_container -v /path/to/data/dir:/data -w /home/ubuntu arkhadem95/dx100 bash
\end{lstlisting}

\textit{Note:} \texttt{/path/to/data/dir} should point to your data directory, which must have at least 20GB of available disk space.

\subsubsection{How to access}

Clone the artifact from our GitHub repository using the following command.

\begin{lstlisting}
git clone https://github.com/arkhadem/DX100.git
cd DX100
export GEM5_HOME=$(pwd)
\end{lstlisting}

\subsubsection{Hardware Dependencies}
This artifact requires 6GB of disk space for simulation infrastructure and benchmarks (\texttt{\$GEM5\_HOME}), and 20GB of disk space for storing simulation results (\texttt{/path/to/data}). The artifact includes 24 simulations (12 benchmarks $\times$ \maa{} and baseline). Each simulation requires approximately 35GB of DRAM memory. Running all simulations on a single core takes about 84 hours to complete. These simulations can be executed in parallel to reduce total runtime, which will increase the resource requirements accordingly. For example, using four cores, the total runtime is reduced to around 24 hours, requiring approximately 140GB of memory.

\subsubsection{Software Dependencies}

If you choose to run the artifact locally instead of using Docker, please ensure the following software dependencies are installed: \texttt{g++-12}, \texttt{clang++-15}, \texttt{SCons >= 3.0}, \texttt{Python >= 3.6}, \texttt{protobuf >= 2.1}, and \texttt{Boost}..

\subsection{Installation}

Building the artifact requires approximately 6GB of disk space and takes around 35 minutes to complete. Use the following script to build all components:

\begin{lstlisting}
# Build Ramulator2
cd $GEM5_HOME/ext/ramulator2/ramulator2/
mkdir build; cd build;
cmake .. -DCMAKE_C_COMPILER=gcc-12 -DCMAKE_CXX_COMPILER=g++-12
make -j

# Build M5ops
cd $GEM5_HOME/util/m5
scons build/x86/out/m5 -j8

#Build gem5
cd $GEM5_HOME
bash scripts/make.sh
bash scripts/make_fast.sh

# Build Benchmarks
cd $GEM5_HOME/benchmarks
bash build.sh
\end{lstlisting}

\subsection{Experiment workflow}

Before running the artifact, you can remove the current results.

\begin{lstlisting}
rm $GEM5_HOME/results
\end{lstlisting}

Use the following script to run the simulations, parse the results, and plot the charts.

\begin{lstlisting}
cd $GEM5_HOME
python3 scripts/benchmark.py -j NUM_THREADS -a all -dir /path/to/data/dir
\end{lstlisting}

Note the following configurations:

\begin{itemize}
  \item \texttt{NUM\_THREADS} specifies the number of parallel simulations. Each simulation requires approximately 35GB of memory, so set this value based on your system's available DRAM.

  \item \texttt{/path/to/data/dir} is the directory where gem5 simulation results will be saved. Ensure it has at least 20GB of free disk space. \textbf{Note:} If you are using the Docker container, set this path to \texttt{/data}.

  \item \texttt{all} runs the full pipeline, including simulation, result parsing, and figure plotting. Alternatively, you can specify \texttt{simulate}, \texttt{parse}, or \texttt{plot} to execute each step individually.
\end{itemize}

\textbf{How Long Simulation Takes?}
Using a single thread, the complete end-to-end execution takes approximately 84 hours. By setting \texttt{NUM\_THREADS} to 4, the execution time is reduced to around 24 hours that requires approximately 140GB of memory.

\textbf{How to Ensure Each Step Runs Correctly?}
You can verify the simulation step by checking the logs located in the \texttt{/path/to/data/} \texttt{dir/results} directory.
After parsing, the raw results will be available in the \texttt{results/results.csv} file.
After plotting, you can find \texttt{png} charts in the \texttt{results} directory.

\subsection{Evaluation and expected results}

You can find the expected raw results (\href{https://github.com/arkhadem/DX100/blob/main/results/results.csv}{\blue{\texttt{results.csv}}}) as well as generated plots corresponding to Figures~\ref{fig:speedup_results}, \ref{fig:bandwidth_results}, and \ref{fig:instr_mpki_results} in the \href{https://github.com/arkhadem/DX100/tree/main/results}{\blue{\texttt{results} directory of the GitHub repository}}.

\end{document}